\documentclass[apj]{emulateapj}

\newcommand{\ngc}[1]{NGC\,#1}

\newcommand{\frms}{$\mathrm{F_{RMS}(\lambda)}$}

\newcommand{\fblue}{$\mathrm{F_{Blue}}$}
\newcommand{\fred}{$\mathrm{F_{Red}}$}

\newcommand{\fdp}{$\mathrm{F_{DP}}$}
\newcommand{\fcbc}{$\mathrm{F_{CBC}}$}
\newcommand{\fratio}{$\mathrm{F_{Red}/F_{blue}}$}

\newcommand{\erg}{erg\,s$^{-1}$\,cm$^{-2}$}
\newcommand{\kms}{km\,s$^{-1}$}
\newcommand{\msigma}{$M_{\bullet} - \sigma_{\star}$}
\newcommand{\msun}{M$_{\odot}$}
\newcommand{\lbol}{$L_{\mathrm{bol}}$}

\newcommand{\mbh}[2]{$\mathrm{M_{\bullet} = #1\,\times\,10^{#2}\,M_{\odot}}$}

\usepackage{float}
\usepackage{amsmath,amstext,amssymb}

\usepackage[usenames,dvipsnames,svgnames]{xcolor}
\usepackage[hyperfootnotes=false, linktocpage=true, colorlinks,linkcolor=Red, citecolor=blue]{hyperref}
\usepackage[textwidth=2.7cm,shadow]{todonotes} 
\usepackage{color, soul}	
\usepackage{verbatim}	

\usepackage{graphicx}

\slugcomment{The Astrophysical Journal, Version of February 2015}
\shorttitle{Variability of double-peaked lines in \ngc{7213}}
\shortauthors{Schimoia et al.}
\begin{document}

\title{Evolution of the accretion disk around the supermassive black hole of \ngc{7213}}


\author{Jaderson S.\ Schimoia, Thaisa Storchi-Bergmann}
\email{silva.schimoia@ufrgs.br}
\affil{Instituto de F\'isica, Universidade Federal do Rio Grande do Sul, Campus do Vale, Porto Alegre, RS, Brazil}

\author{Cl\'audia Winge}
\affil{Gemini South Observatory, c/o AURA Inc., Casilla 603, La Serena, Chile}
\author{Rodrigo S. Nemmen}
\affil{Universidade de S\~ao Paulo, Instituto de Astronomia, Geof\'{\i}sica e Ci\^encias Atmosf\'ericas, Departamento de Astronomia, \\ S\~ao Paulo, SP 05508-090, Brazil}
\and
\author{Michael Eracleous}
\affil{Department of Astronomy and Astrophysics and Institute for Gravitation and the Cosmos, Pennsylvania State University, \\525 Davey Lab, University Park, PA 16802, USA}

\begin{abstract}

We present observations of the double-peaked broad H$\alpha$ profile emitted by the active nucleus 
of NGC\,7213 using the the Gemini South Telescope in 13 epochs between 2011 September 27 and
2013 July 23. This is the first time that the double-peaked line profile of this nucleus -- typical of gas emission
from the outer parts of an accretion disk  surrounding a supermassive black hole (SMBH) --  is reported to vary.
From the analysis of the line profiles we find two variability timescales: (1) the shortest one, between 7 and 28 days, 
is consistent with the light travel time between the ionizing source and the part of the disk emitting the line; and (2) a 
longer one of $\gtrsim 3$ months corresponding to variations in the relative intensity of the blue and red sides of the profile, 
which can be identified with the dynamical timescale of this outer part of the the accretion disk. We modeled the line profiles as due to emission from a 
region between $\approx$ 300 and 3000 gravitational radii of a relativistic, Keplerian accretion disk  surrounding the SMBH.
Superposed on the disk emissivity, the model includes an asymmetric feature in 
the shape of a spiral arm with a rotation period of $\approx$21 months, which reproduces the variations in the relative 
intensity of the blue and red sides of the profile. Besides these variations, the $rms$ variation profile reveals the presence of 
another variable component in the broad line, with smaller velocity width W$_{68}$ (the width of the profile corresponding to 68$\%$ of the flux) of $\sim 2100$\,\kms.
\end{abstract}

\keywords{accretion, accretion disks --- galaxies: individual (NGC\,7213) --- galaxies: nuclei --- galaxies: Seyfert --- line: profiles}

%
%
%
\section{Introduction}

NGC\,7213 is a nearby (z=0.005839) Sa spiral galaxy. Its active nucleus was first classified as 
Seyfert\,1 by \citet{Phillips79} based on its very broad H$\alpha$ emission line 
(13,000 \kms\ for the full width at zero intensity) and later was also recognized as
low-ionization nuclear emission-line region (LINER) by \citet{Filippenko84}
based on a study of a variety of optical emission lines which were observed to have full width at half maximum (FWHM)  
in the range of 200 -- 2000 \kms. 
Its nucleus harbors a $\sim 10^8$ \,\msun \ supermassive black hole (SMBH) \citep{Woo02}, producing a bolometric luminosity of \lbol\,=\,7$\times10^{43}$erg\,s$^{-1}$ \citep{Emmanoulopoulos12}.

The nucleus of NGC\,7213 has been extensivelly studied in X-rays. XMM-Newton/BeppoSAX 
observations revealed that the AGN spectrum shows no significant Compton reflection component \citep{Bianchi03}
-- what is very peculiar among bright Seyfert\,1 AGN's. Additionally,  these observations also revealed 
 the presence of a significant Fe\,K complex. \citet{Bianchi08} reported the
data analysis of a long Chandra HETG observation
finding that the neutral iron K$\alpha$ line has a FWHM of 2400$^{+1100}_{-600}$\,\kms, 
claiming that it was fully consistent with the H$\alpha$ FWHM (2640$^{+1100}_{-600}$\,\kms).
This seems to be at odds with the previous studies that showed a much broader profile or
it may indicate a strong variation of the profile.
\citet{Bianchi08} concluded that the neutral Fe\,K line originates in the Compton-thin Broad Line Region (BLR) 
to explain the absence of Compton reflection and of any relativistic component in the lines.
More recently, \citet{Lobban10} reproduced the highly ionized iron K lines
with a photoionization model where gas has a column density of $N_H > 10^{23}$\,cm$^{-3}$ and is 
likely to be located between 10$^3$ -- 10$^4$ gravitational radii ($R_{g}$) from the central source. 
The authors suggest that the inner accretion flow is radiatively inefficient \citep[RIAF, e.g.,][]{Yuan14}, 
which would explain the photoionization spectrum and the absence of the optically thick disk component.

Although the broad H$\alpha$ emission line of NGC\,7213 has been reported in previous works
\citep{Phillips79, Filippenko84, SB96}, there are no variability studies of its profile to date.
A recent study by \citet{Allan14} (see Figure 1) showed its 
H$\alpha$ line to present a very broad double-peaked component.
Very broad double-peaked lines -- with velocity separation of $\sim$10,000 \kms\, between the blue and red peaks are thought to originate 
in the outer parts of an accretion disk \citep{Schimoia12, Lewis10}.
Models of ionized gas rotating in a relativistic Keplerian accretion
disk around a SMBH have been successful in accounting for such
double-peaked profiles \citep{Chen89, CeH89, SB03, Strateva03, Lewis10}. 
These models explain most of the observed features of the
profiles \citep{Eracleous03} and can help to constrain
the physical properties of the line emititing part of the disk, as, for instance, the inner and outer
limits for the emitting region as well as its inclination relative to the plane of the sky.
Although other models have been considered for the origin of the double-peaked profiles,
these are much less attractive than the accretion disk model (see discussion in \citet{Eracleous03, Eracleous09})

Double-peaked profiles are expected to vary.
 \citet{Lewis10} and \citet{Gezari07} studied such variation for 14 double-peaked emitters in  
a long-term monitoring of these sources at  time intervals ranging from several months to years.
They found that all profiles showed variability on timescales of years. 
Recently, \citet{Schimoia12, Schimoia15},
monitored the double-peaked H$\alpha$ profile of NGC\,1097
on shorter timescales and found that, besides presenting variations on similarly long (1.5 years) timescales,
the profile also varied on timescales as short as a week or even shorter.
They concluded that there are different timescales of variability in this source, 
which should apply also to other double-peak emitters, including NGC\,7213.

In the present paper we model, for the first time, the double-peaked H$\alpha$ profile of 
the NGC\,7213 nucleus
as due to emission from the outer parts of a relativistic, Keplerian accretion disk,
constraining its properties. We also present a study of its variation from
spectral monitoring of the profile over a time span
of almost two years, from Sep 27, 2011 to Jul 23, 2013, 
including one time interval as short as a week. We find that the variation of the profiles reveal
that, not only the double-peaked profile varies, on short and long timescales, but there is another 
broad component, with a velocity width\,$\sim$\,2100\,\kms, that is also variable.

This paper is organized as follows: in \S 2 we describe the
observations and the data reduction; 
in \S 3 we present the observational results, the adopted accretion disk model, and 
the discovery of an additional variable component to the profile; in
\S 4 we discuss the timescales of the accretion disk variability, the
interpretation and implications of the modeling to the structure of the AGN and explore 
a determination of the mass of the SMBH through the model. The conclusions of this work are
presented in \S 5.

\section{Observations and data reduction}
We obtained a total of 13 optical spectra of the nucleus of the galaxy NGC\,7213 from 2011 September 27 to 2013 July 23. 
The observation of 2011 September 27 was obtained with the Integral Field Unit of
the Gemini Multi Object Spectrograph (GMOS-IFU) at the Gemini South telescope
\citep[Gemini project GS-2011B-Q-23]{Allan14}.
These observations consisted of two adjacent IFU fields (covering  7$\arcsec$\,$\times$ 5$\arcsec$ each)  
resulting  in  a  total  angular coverage of 7$\arcsec$\,$\times$\,10$\arcsec$ around the nucleus.
The wavelength range of this observation was 
5600-7000 \AA\ in order to cover  H$\alpha$+[N II] $\lambda\lambda$6548,6583 \AA\
and [S II] $\lambda\lambda$6716,6731 \AA\ observed with the grating R400-G5325, which resulted
in a resolution of R $\approx$ 2000 ($\sim$\,150\,\kms).
The seeing during the IFU obervations was 0$\farcs$5, what results in a spatial resolution of 58\,pc, and the
sampling of the final reduced cube is 0$\farcs$1$\times$0$\farcs$1.

The remaining 12 observations were taken between 2012 July 21 and 2013 July 23 with the spectrograph GMOS of the 
Gemini South telescope in the longslit mode (project ID GS-2012A-Q-86). The slit used was 1$\farcs$0 wide
and 330\,$\arcsec$ long and was oriented at the position angle of 305$^{\circ}$ in all observations. The grating was the B600-G5323
with the central wavelength of 5700 \AA\ chosen to cover the H$\alpha$+[N II] $\lambda\lambda$6548,6583 \AA\
and H$\beta$ $\lambda$4862 \AA\ and give a spectral resolution of R\,$\approx$\,1688 ($\sim$177\,\kms).
The instrumental setup of the longslit observations resulted in a pixel scale of 0.14 $\arcsec$/pixel.
Most visits consisted of 6 exposures each, giving a total of 2700s on source.
Table \ref{n7213_obs} lists the dates of the observations, the instruments, number and length of exposures
on each visit. The data were reduced
using the standard procedures and packages for IFU and longslit modes in
\textit{IRAF}\footnote{IRAF is distributed by the National Optical Astronomy Observatories, 
which are operated by the Association of Universities for Research 
in Astronomy, Inc., under cooperative agreement with the National 
Science Foundation.}. Throughout the paper, in all figures, the spectra are shown at the rest wavelength.

\begin{center}
\begin{deluxetable}{l c c c }
\tablecolumns{4}
\tablecaption{Observations of NGC\,7213}
\tablehead{UT Date	&MJD 	&Mode &Exposures}
\startdata 

Sep 27 2011 &55831.130 &IFU      &12$\times$350 \\
Jul 21 2012 &56129.155 &Longslit &6$\times$450 \\
Jul 30 2012 &56138.422 &Longslit &6$\times$450 \\
Oct 15 2012 &56215.111 &Longslit &6$\times$450 \\
Nov 22 2012 &56253.047 &Longslit &6$\times$450 \\
Apr 13 2013 &56395.383 &Longslit &6$\times$450 \\
May 11 2013 &56423.386 &Longslit &6$\times$450 \\
May 20 2013 &56432.315 &Longslit &6$\times$450 \\
May 30 2013 &56442.334 &Longslit &6$\times$450 \\
Jun 14 2013 &56457.399 &Longslit &6$\times$450 \\
Jun 30 2013 &56473.310 &Longslit &6$\times$450 \\
Jul 07 2013 &56480.218 &Longslit &6$\times$450 \\
Jul 23 2013 &56496.278 &Longslit &6$\times$450 
\label{n7213_obs}
\tablecomments{Column (1) gives the date of observations while column (2) gives the Modified Julian Date (JD$-2400000.5$).
Column (3) is the mode of observation and the column (4) gives the number of visits and the exposure time of each visit.}
\end{deluxetable}
\end{center}

\subsection{Subtraction of the underlying stellar population contribution}

\begin{figure}
\centering
\includegraphics[width=8cm]{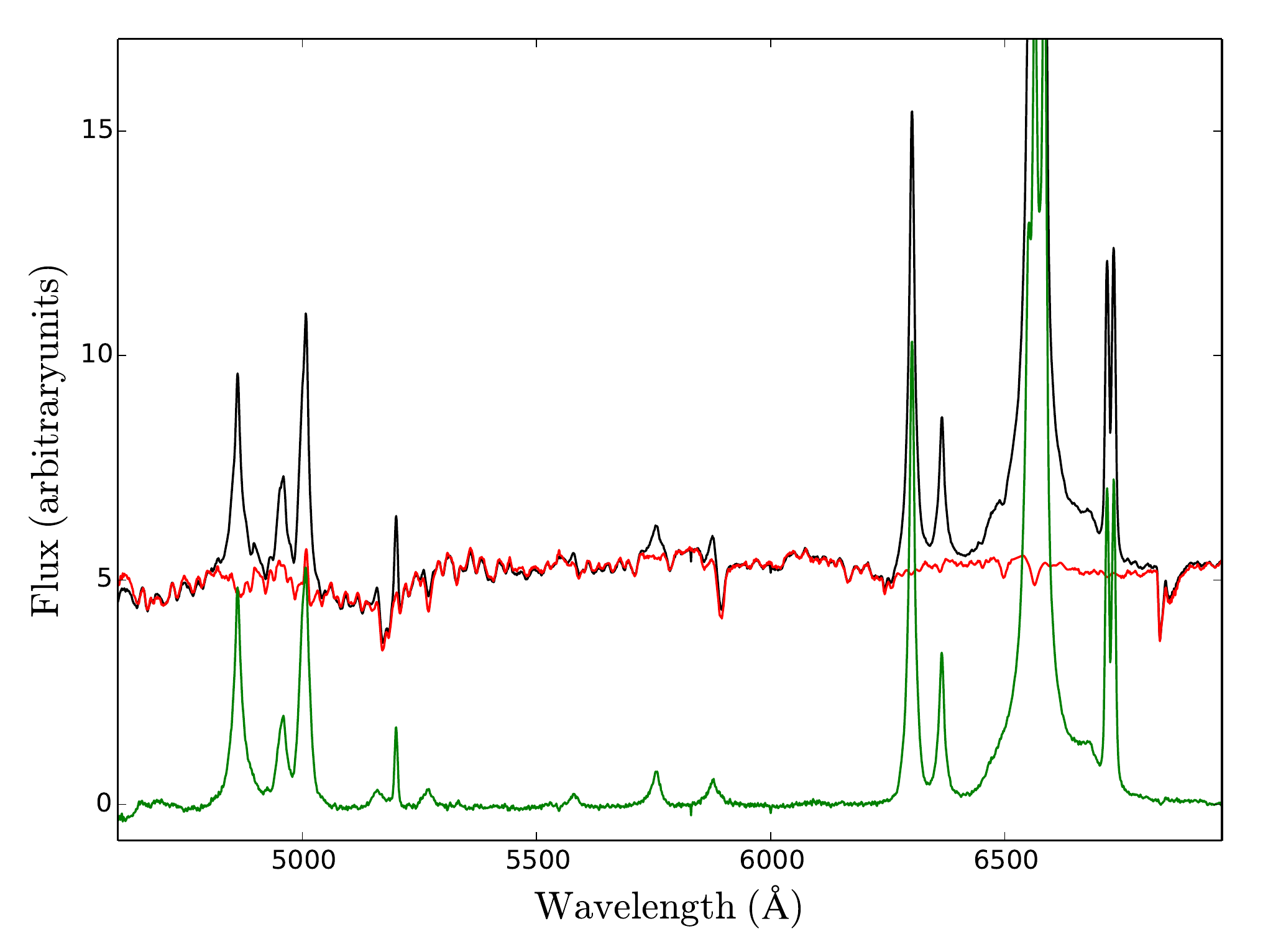}
\caption{Example of the process of stellar population subtraction for the spectrum obtained in 2012 July 21. \emph{Top}: the black solid line is the nuclear spectrum extracted within a window of 1$\farcs$0$\times$1$\farcs$0; the red solid line is the stellar population template obtained by averaging two extranuclear spectra at 2$\farcs$0  from the nucleus, scaled to match the continuum of the nuclear extraction. \emph{Bottom}: the green solid line is the resulting \emph{nuclear emission spectrum}, after subtracting the stellar population template.
}
\label{stellar_pop}
\end{figure}

We extracted the nuclear spectra from our data using a window of 1$\farcs$0$\times$1$\farcs$0 centered
on the peak of the continuum emission. Within this aperture,
the continuum is dominated by the underlying stellar population. 
In order to isolate the AGN emission
we subtracted the contribution of the stellar population. The process is illustrated in Figure \ref{stellar_pop}, 
and is described as follows:
\begin{itemize}
 \item First we extracted the nuclear spectrum;
 \item Two extranuclear spectra, also within  windows of 1$\farcs$0$\times$1$\farcs$0, were extracted at 2$\farcs$0 
 away from both sides of the nucleus;
 \item The two extranuclear spectra were averaged. The average spectrum was scaled to match the flux of the continuum of the nuclear spectrum, as it displays the same absorption features with the same equivalent widths to those of the nuclear spectrum. This indicates that the nuclear spectrum shows no detectable non-stellar continuum emission, only line emission (as in our previous studies of NGC\,1097 \citet{SB03, Schimoia15}).
The average extranuclear spectrum is only weakly ``contaminated" by narrow emission lines which were excised 
by using a synthetic spectrum obtained from the application of the {\tt starlight-v04} spectral synthesis code of \citet{Cid};
 \item The scaled and corrected stellar population template was then subtracted from the nuclear spectrum in order to 
 isolate the AGN emission.
 
\end{itemize}

After the subtraction of the stellar population template we adopted the spectrum of 2011  September 27 (MJD\,55831) -- which has the best signal-to-noise ratio (12$\times$350\,s of exposure on the source) -- as the reference spectrum,
and calibrated the other spectra by scaling the flux of the narrow lines 
[O\,I]$\lambda$6300\,\AA\ and [O\,I]$\lambda$6363\,\AA\ to match those of the reference spectrum.
The [O\,I] lines are thought to originate in a larger region than that producing
the broad double-peaked Balmer lines, thus these [O\,I] lines should not display significant variations in their flux during the time span of the observations, 
these lines are also isolated and unaffected by broad, variable emission lines.
After this ``intercalibration" by the [O\,I] emission lines we measured the flux of the narrow [S II] λλ6716,6731\,\AA\
lines in the different spectra finding small variations of 7 -- 8\% in the line fluxes. We thus conclude that the flux calibration among the spectra leads to uncertainties of 7 -- 8 \%\ in the line fluxes.
This uncertainty was added in quadrature to the other uncertainties in our measurements.

\section{Results}
\label{results}
\begin{figure*}
\centering
\includegraphics[width=17cm]{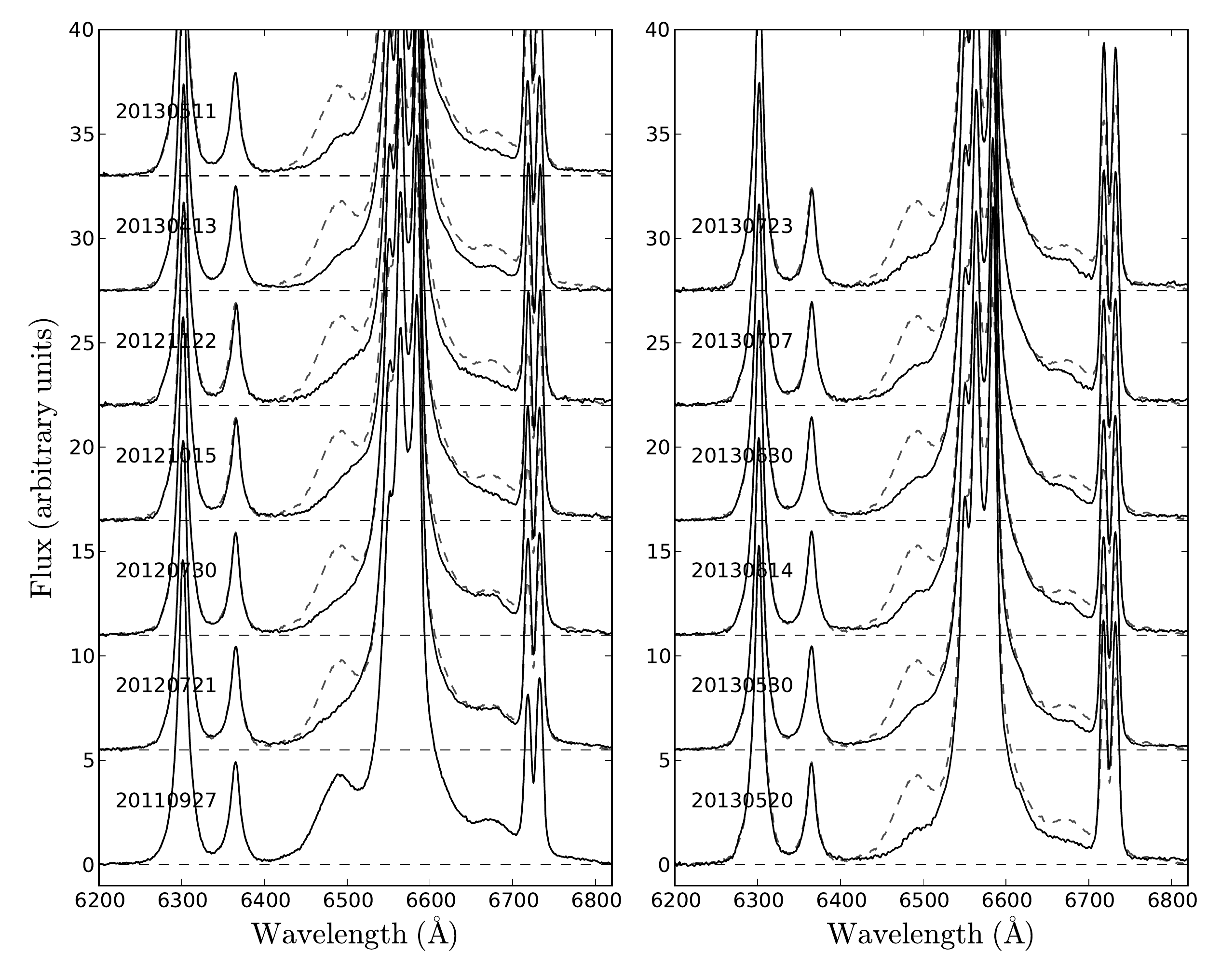}
\caption{The nuclear emission line profiles from 2011 September 27 (MJD 55831) 
to 2013 July 23 (MJD 56496). The reference spectrum of 2011 September 27 (MJD\,55831) is displayed in each panel in  dashed lines for comparison.}
\label{spectra}
\end{figure*}

The resulting nuclear emission spectra, after subtraction of the underlying stellar population and calibration by the narrow emission lines, are shown in Figure \ref{spectra}.
This figure shows that in the first observation of 2011 September 27 the blue and red peaks were very prominent in the H$\alpha$ profile, with the blue peak higher than the red peak. During the second observation of 2012 July 21 (MJD\,56129), taken almost ten months after the first, the profile shows less pronounced peaks, mainly the blue peak, whose flux decreased to the point of becoming lower than the red peak.
The red side of the profile is now more prominent than the blue side. 
The overall flux of the double-peaked profile decreased from 2011 September 27 to 2012 July 21, 
thus showing that the time interval of ten months between these two observations is large 
enough to allow significant changes in the shape and flux of broad H$\alpha$ double-peaked profile.

After this long time interval between the first two observations, we monitored the profile more frequently
in order to look for shorter variability timescales. We found such variations. For instance, from 2012 Jul 30 (MJD\,56138) to 2012 October 15 (MJD\,56215), the profile changed showing mainly a decrease in flux in the red wing. In the subsequent spectra the profile continued to change, with the overall flux always lower than that of the first observation. While in the first observation one can clearly see a blue and a red peak
in the profile, in most of the other observations the shape of the profile can be better 
described as showing a double shoulder instead of double peak.


\subsection{Measurements of the flux variations of the broad double-peaked emission line}
\label{measurements}

\begin{figure}
\centering
\includegraphics[width=8cm]{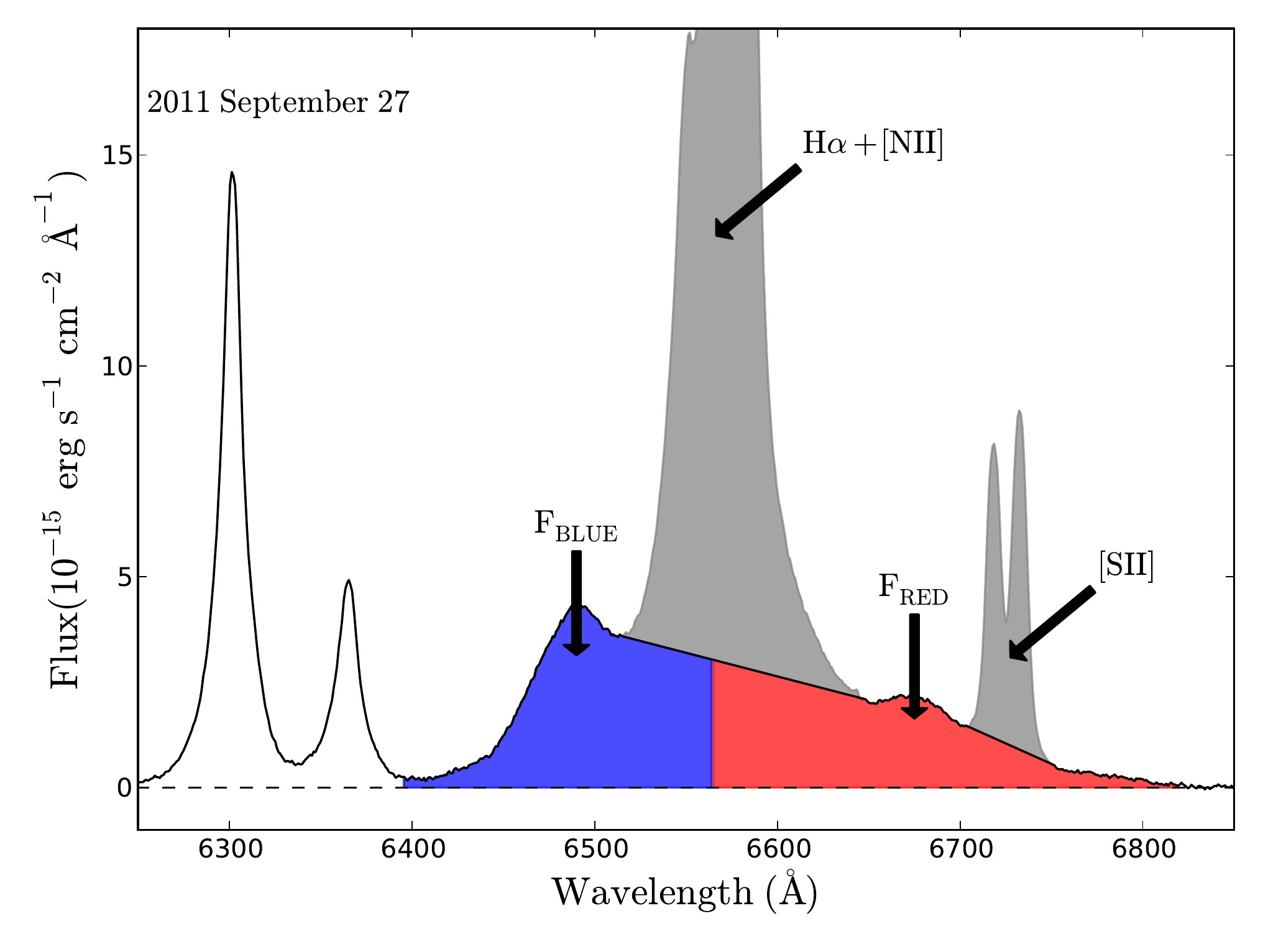}
\caption{The measured characteristics of the broad double-peaked H$\alpha$ profile \fblue\ and \fred. 
The gray regions are emission lines that were removed before measuring the flux of the double-peaked profile.
The blue region, between 6395\AA\ and 6563\AA\, shows the area integrated to obtain the flux of the blue side of the profile, \fblue.
The red region, between 6563\AA\ and 6820\AA\, shows the same for the red side of the profile, \fred.}
\label{fbroad_measure}
\end{figure}

In order to measure the flux of the double-peaked line and quantify its variations, we proceeded as follows:

\begin{itemize}
\item We first removed the contribution of the narrow emission lines interpolating a linear effective continuum between $\sim$6513\,\AA\ and $\sim$6648\,\AA\ excising the emission lines H$\alpha$+[N II]$\lambda\lambda$6548,6583\,\AA\ above this continuum. We performed the same
process between $\sim$6703\,\AA\ and $\sim$6751\,\AA\ to remove the [S II] $\lambda\lambda$6716,6731\,\AA\ emission lines. The removed emission lines are shown as gray regions in Figure \ref{fbroad_measure}.
 
 \item After the removal of the narrow emission lines we defined two parameters: \fblue\, 
 which is the integrated flux under the blue side of the broad double-peaked profile, 
 between 6395\AA\ and 6563\AA\ and \fred\, which is the  integrated flux under the red side of 
 the double-peaked profile, between 6563\AA\ and 6820\AA. Since we have corrected the spectra by the
 redshift, the definition of 6563 \AA\, as the wavelength that divides the fluxes of the blue and red sides
 is equivalent to use as reference the systemic velocity of the galaxy, as
 6563 \AA\, is the rest wavelength of H$\alpha$.
 
 \fblue\ is illustrated as the blue  
 region in Figure \ref{fbroad_measure} while \fred\ is the red  region in the same figure. 
We note that our definition is such that the wavelength range covered by the red side
is wider than that of the blue side of the profile. This effect is expected due to 
(at least partially) the gravitational redshift, since the high velocity wings of the 
profile come from gas that is very close to the SMBH ($\sim$ 300 gravitational radius;  
see following discussion on the modelling of the profile).
 


 \item \fdp\ was then obtained as the sum of \fblue\ and \fred\ (being thus the total integrated flux of the double-peaked line), while \fratio\ was obtained from the ratio between \fblue\ and \fred.
\end{itemize}

\begin{figure}
\centering
\includegraphics[width=8cm]{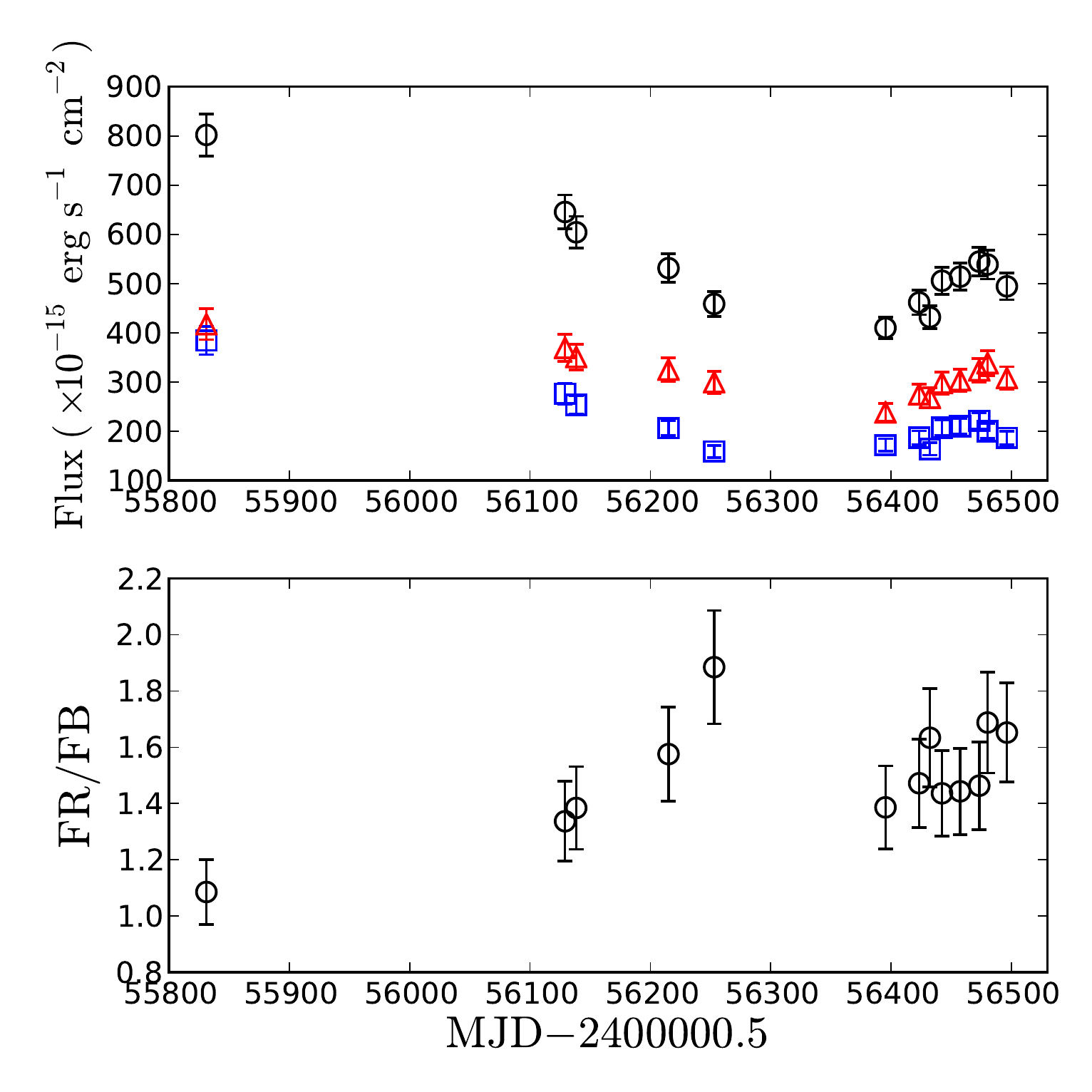}
\caption{\emph{Top panel}: The black circles represent the integrated flux of the broad double-peaked line, \fdp. The blue squares represent the integrated flux of the blue side of the line -- \fblue, while the red triangles represent the 
integrated flux of the red side of the line -- \fred. All measurements were performed 
after excising the narrow emission lines as illustrated in Figure \ref{fbroad_measure}. \emph{Bottom panel}: the black circles represent the temporal evolution of \fratio -- the ratio between \fred\ and \fblue. 
}
\label{light_curve}
\end{figure}

The parameters defined above are similar to those used to quantify the variability of the double-peaked profile of NGC\,1097 \citep{Schimoia15}. The main difference is that the double-peaked profile of NGC\,7213, in some epochs, does not display clear blue and red peaks, but shoulders, instead. Thus,
instead of measuring the flux density and position of the blue and red peaks we rather measured the 
integrated flux of each side of the double-peaked line, which is more robust for this case.
The values of \fblue, \fred, \fdp\ and \fratio\ are listed in Table \ref{fbroad_data}, while the time variation of these parameters is shown in Figure \ref{light_curve}.

\subsection{Variability Timescales}

From our previous studies of the broad double-peaked H$\alpha$ profile of NGC\,1097
we concluded that it shows two main variability timescales: a long one, identified as a dynamical timescale for the gas orbiting in the disk (see \S 4); and a shorter one that can be identified
with the light travel time between the ionizing source and the emitting portion of the disk. In NGC\,7213, we tentatively identify these timescales as follows:

\begin{itemize}
 \item \emph{Long variability timescale}: significant changes in the integrated flux of the double-peaked profile 
 can occur on timescales of a few months. For instance, from 2012 July 21 (MJD\,56129) to 2012 November 22 (MJD\,56253)  \fdp\ 
 decreased by $\sim$30\% in approximately 4 months. During the same 
 period, the relative intensity of the red and blue sides of the profile, \fratio, also changed by 46\%.
But we also note that between 2013 April 13 (MJD\,56395) and 2013 July 23 (MJD\,56496) -- a time interval of 
 $\sim$ 100 days -- \fratio\ almost did not vary, keeping a value of $\sim$ 1.5. We thus conclude that the timescale of the  \fratio\ variations is $\gtrsim$ 3 months. 
 
 \item \emph{Short variability timescale}: from 2013 April 13 to 2013 July 23 we obtained more frequent observations, 
 separated by time intervals of a week to a month. The shortest time interval between consecutive observations is 7 days (from 2013 June 30 to 2013 July 07). 
 On this timescale \fdp\ did not display changes larger than the uncertainties in the measurements. During the period from 2013 April 13 (MJD\,56395) to 2013 May 11 (MJD\,56423) \fdp\ changed from 409.8$\pm$22.1 to 461.6$\pm$25\,$\times10^{-15}$\erg\, thus by 10\% in a time interval of 28 days. We can thus only put limits on the shortest timescale: it is probably longer than a week but shorter than 28 days. We note that this week-to-28 days for the shortest timescale of variability was obtained for our specific set of observations. Since we did not sample the profile in a time interval shorter than 7 days we may have missed possible stochastic variations that may have occurred on a shorter timescale.

\end{itemize}

\begin{center}
\begin{deluxetable*}{l c c c c c }
\tablecolumns{6}
\tablecaption{\ngc{7213} double-peaked line measurements}
\tablehead{UT Date		&MJD 	&\fblue 		 &\fred   	    &\fdp &\fratio}
\startdata                                                            
Sep 27 2011 &55831.130 &384$\pm$29 &417$\pm$31 &802$\pm$43 &1.1$\pm$0.1 \\ 
Jul 21 2012 &56129.155 &276$\pm$21 &369$\pm$28 &646$\pm$35 &1.3$\pm$0.1 \\ 
Jul 30 2012 &56138.422 &253$\pm$19 &351$\pm$26 &604$\pm$33 &1.4$\pm$0.2 \\ 
Oct 15 2012 &56215.111 &206$\pm$16 &325$\pm$24 &532$\pm$29 &1.6$\pm$0.2 \\ 
Nov 22 2012 &56253.047 &159$\pm$12 &300$\pm$23 &458$\pm$26 &1.9$\pm$0.2 \\ 
Apr 13 2013 &56395.383 &173$\pm$13 &238$\pm$18 &410$\pm$22 &1.4$\pm$0.2 \\ 
May 11 2013 &56423.386 &187$\pm$14 &275$\pm$21 &462$\pm$25 &1.5$\pm$0.2 \\ 
May 20 2013 &56432.315 &164$\pm$12 &268$\pm$20 &432$\pm$24 &1.6$\pm$0.2 \\ 
May 30 2013 &56442.334 &208$\pm$16 &298$\pm$22 &506$\pm$27 &1.4$\pm$0.2 \\ 
Jun 14 2013 &56457.399 &210$\pm$16 &304$\pm$23 &514$\pm$28 &1.5$\pm$0.2 \\ 
Jun 30 2013 &56473.310 &221$\pm$17 &324$\pm$24 &545$\pm$29 &1.5$\pm$0.2 \\ 
Jul 07 2013 &56480.218 &200$\pm$15 &338$\pm$25 &539$\pm$30 &1.7$\pm$0.2 \\ 
Jul 23 2013 &56496.278 &187$\pm$14 &308$\pm$23 &495$\pm$27 &1.7$\pm$0.2 \\ 

\label{fbroad_data}
\tablecomments{ Column (1) gives the date of observations while column (2) gives the Modified Julian Date (JD$-2400000.5$).
Column (3) gives the integrated flux of the blue side of the double-peaked line in 
units of $10^{-15}$\,erg\,s$^{-1}$\,cm$^{-2}$ and the column (4) gives the integrated flux of the red side of the line in the same units. Column (5) gives total flux of the double-peaked line.
Column (6) gives the ratio between (4) and (3).}
\end{deluxetable*}
\end{center}



\subsection{The $rms$ spectrum}

Figure \ref{spectra} shows that the broad double-peaked profile clearly varies with time. 
Variations in the illumination of the disk and/or the presence of structures eventually developing in the disk can explain these variations. In order to investigate which regions of the profile show more variations, we have 
calculated the \emph{root mean square} ($rms$) spectrum -- \frms\ -- which can reveal if 
there are specific regions in the velocity space in which the flux varies more. 
The $rms$ spectrum, $\mathrm{F_{RMS}(\lambda)}$, 
is calculated as the $rms$ variation of the flux at each wavelength covered by the double-peaked profile.
As discussed in \S \ref{measurements}, significant changes in \fdp\ and in the shape of the double-peaked profile 
are only seen on timescales $\gtrsim$ 30 days. Thus, in order to calculate the $rms$ spectrum, we selected spectra that are separated by time intervals of the order of 30 days, namely:
2011 September 27, 2012 July 21, 2012 October 15, 2012 November 22, 2013 April 13, 2013 May 20, 2013 June 14, 2013 July 07.

\begin{figure}
\centering
\includegraphics[width=8cm]{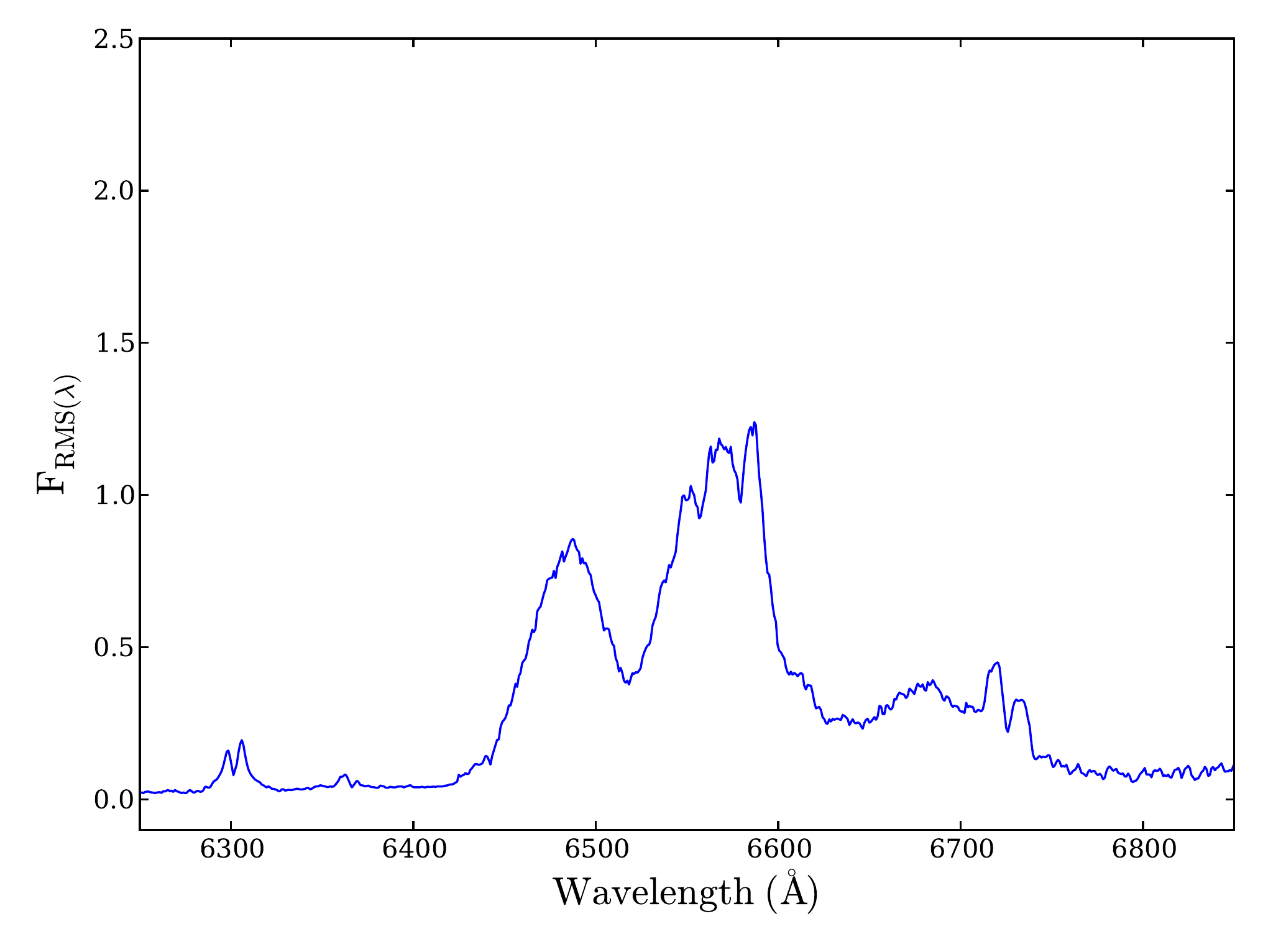}
\caption{The $rms$ scpectrum calculated from our data. The blue and red humps are observed at the wavelengths of the two peaks of the double-peaked profile, but the central hump shows the presence of another  component showing similar variation to that of the double-peaded component. Some residuals, of the order ot the uncertainties in the cross-calibration of the spectra, are observed at the wavelengths of the narrow emission lines.}
\label{rms_only}
\end{figure}

The resulting $rms$ spectrum is shown in Figure \ref{rms_only}. The largest variations in the profile
are observed in three ``peaks": two of them coinciding in wavelength with the blue and red peaks of the double-peaked profile and a third peak observed in between the two peaks, centered at a velocity close to zero (adopted as the velocity of the narrow H$\alpha$ component). The double-peaked structure is expected for variable emission
in an accretion disk, but the central hump reveals that there is significant variation also in a lower velocity structure,
resembling another broad component with a velocity width of $\approx$\,2100\,\kms\ (see discussion below). Some residual variation is also observed at the wavelengths of the narrow lines, but this is due to the estimated uncertainties  
of the ``intercalibration" of the data, as discussed previously.

The blue and red peaks of the $rms$ spectrum support the origin of the double-peaked emission and its variability, from gas emission in an accretion disk with variable surface emissivity.
We will thus model such emission with an Keplerian relativistic accretion disk in the next sections.
But, besides the double-peaked component, there are also 
comparable flux variations at lower projected velocities, indicating the presence of another varying component besides the one attributed to the accretion disk. This component could
be due to emitting clouds orbiting the SMBH at distances beyond the outer border of the disk, as it shows much smaller FWHM than that of the double-peaked line. We will call this component ``central-broad component", or CBC, for short.
In our modelling, described in the next sections, we will take into account the contribution of this component via the fit of three Gaussian components to it.

\subsection{The Accretion Disk Model}
\label{accretion_disk}



%
We modeled the double-peaked H$\alpha$ profiles using the accretion disk model described by
\citet{Gilbert99}, \citet{SB03}, and \citet{Schimoia12, Schimoia15}. In this
formulation, the broad double-peaked emission line originates in a
relativistic Keplerian disk of gas surrounding the SMBH. The
line-emitting portion of the disk is circular and located between an
inner radius $\xi_{1}$ and an outer radius $\xi_{2}$  -- where $\xi$ is
the disk radius in units of the gravitational radius $r_g={GM_{\bullet}}/{c^2}$, $c$ is the speed of light, $G$ is the gravitational constant,
and $M_{\bullet}$ is the mass of the black hole. The disk has an inclination angle $i$ 
with respect to the plane of the sky.

In order to take into account the observed asymmetries on the double-peaked profile we adopted
the ``saturated spiral model'' \citep{Schimoia12} for the total
emissivity of the accretion disk. In this formulation, there is an enhancement of the emissivity 
in the form of a spiral arm, which is superposed on the underlying emissivity of the circular accretion disk.
The details of the emissivity law are described in \citet{SB03} and  \citet{Schimoia12}. Here we briefly describe the physical meaning of the different parameters which are relevant for our modeling.

The parameter $\xi_{q}$ is the radius of maximum emissivity, or saturation
radius, at which the emissivity law changes; $q_{1}$ is the index of
the emissivity law for $\xi_{1} < \xi < \xi_{q}$; $q_{2}$ is the index
for $\xi_{q} < \xi < \xi_{2}$.
The presence of the spiral arm enhances the emissivity of 
of the gas where it is located, thus the parameter $A$ represents
the brightness contrast between the emissivities of the spiral arm and the
underlying disk.
The emissivity of the spiral arm decays as a function of the azimuthal
distance $\phi - \psi_{0}$ from the ridge line  to both sides of the
arm, assumed to be a Gaussian with an azimuthal width of $\delta$. 
Furthermore, $\phi_0$ is the azimuthal angle of the spiral
pattern, $p$ is the pitch angle, and $\xi_{sp}$ is the innermost radius
of the spiral arm (cf. \citet{Schimoia15} for more details).

We kept the emissivity index for radii larger than the break radius as $q_2=3$, as proposed 
by \cite{Dumont90}: after the saturation radius the emissivity of the disk is expected to be proportional 
to $\xi^{-3}$. We tested many values for the emissivity index for the region between the inner
and the saturation radius, $\xi_1<\xi<\xi_q$,
and concluded that $q_1 = -0.2 $ gave the best fits; this small value for $q_1$ is required because,
as can be seen in Figure \ref{spectra}, the broad double-peaked profile displays very extended wings. 
The presence of extended wings means that the inner parts, 
with higher projected velocities, are important for the emissivity of the disk and the value $q_1=-0.2$
implies that the emissivity increases slowly until the break radius,
which makes the inner parts of the accretion disk important with respect to the outer parts.
We also tested different values for the inner and outer radii, 
$\xi_1$ and $\xi_2$, and the inclination angle, $i$, and found that the set of values
that best reproduced all the data together
are $\xi_1 = 300\,\pm\,60$, $\xi_2 = 3000\,\pm\,90$ and $i=47^{\circ}\,\pm\,2^{\circ}$, so
in the analysis below, we keep these values for these parameters. 

\begin{figure}
 \centering
 \includegraphics[width=8cm]{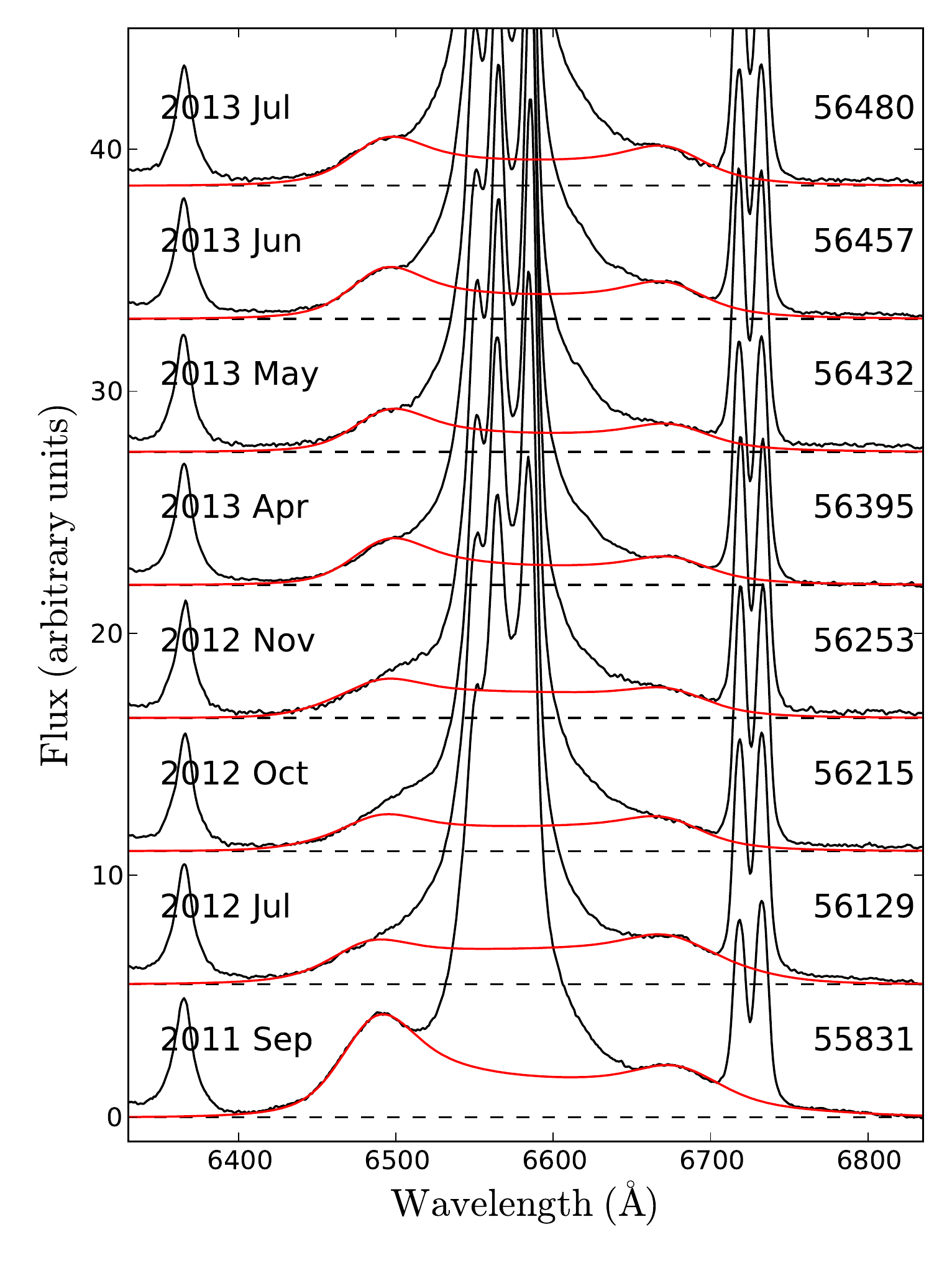}
 \caption{The black solid line is the profile observed in different epochs, after subtraction of the stellar continuum and
calibration by the flux of the narrow lines. In each frame the red solid line represents the emission of the accretion disk  model that best fits the double-peaked profile. The spectra are shifted in flux for better visualization.
 The parameters of each fit are listed in Table \ref{model_pars}.}
 \label{model_fit}
\end{figure}

\begin{figure}
\centering
\includegraphics[width=8cm]{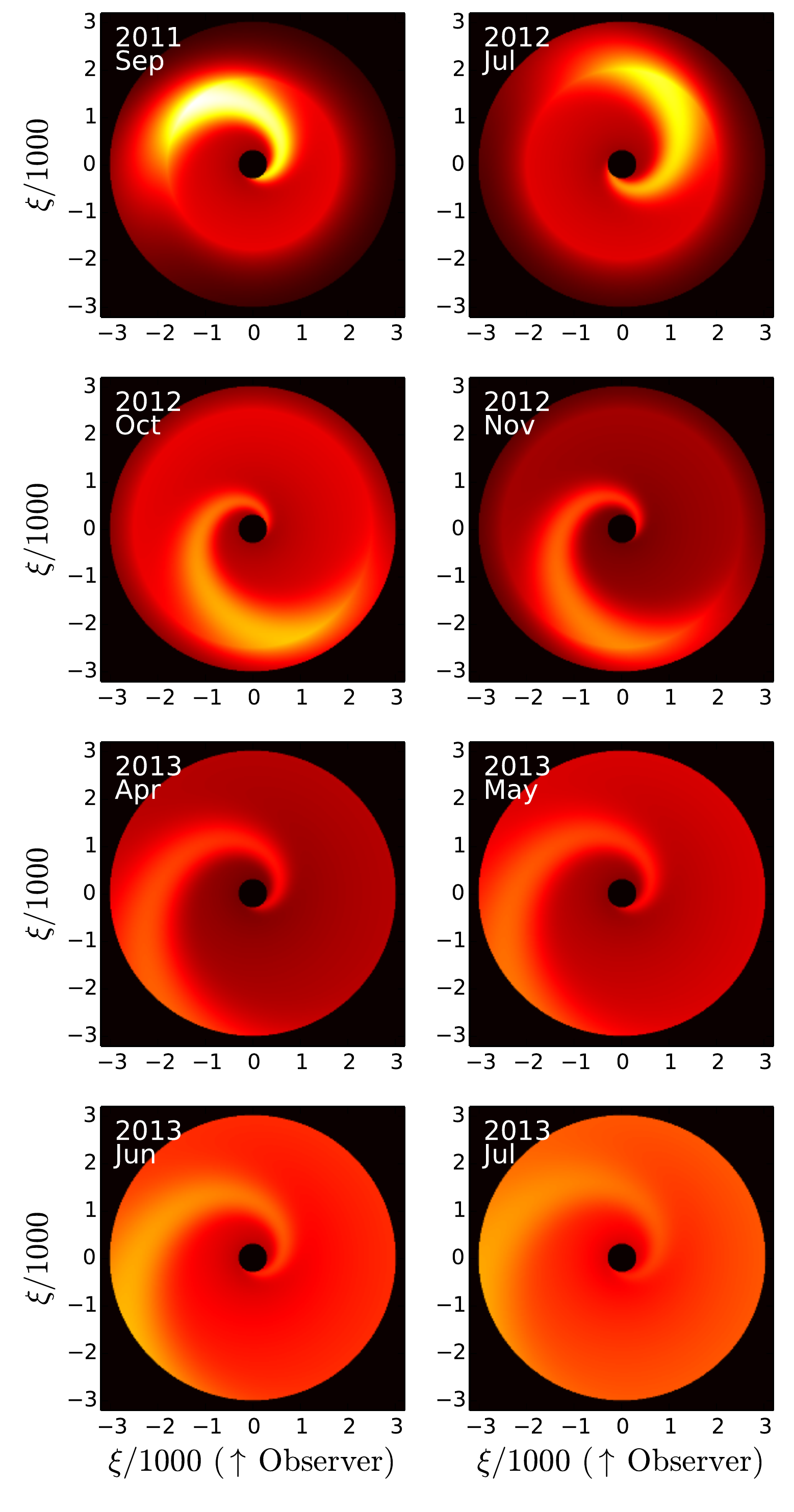}
\caption{Emissivity maps corresponding to the accretion disk models fitted to the double-peaked profiles of Figure \ref{model_fit}. The yellow to white represents regions with the highest surface emissivity while the dark red represents the regions with the lowest surface emissivity.
The non-axisymmetric part of the emissivity has the shape of a spiral arm rotating in the accretion disk. The observer sees the disk from the bottom of the figure and the spiral arm rotates clockwise. The epoch of observation is written in the top left corner of each frame.}
\label{disk_model}
\end{figure}

In order to reproduce the double-peaked profile variations, namely, in the relative intensity of the blue and red sides of the double-peaked profile and the broadening/narrowing of the profile, we first modeled the reference spectrum (with the best signal-to-noise ratio), from  2011 September 27; and after finding the best parameter values for this epoch, we allowed only 3 parameters to vary:
$\phi_0$, $A$ and $\xi_q$. The parameters that define the shape of the spiral arm were kept fixed 
at $p=13^{\circ}$, $\delta=75^{\circ}$ and $\xi_{sp}=\xi_1$. Among these parameters, the pitch angle is the less uncertain because it regulates the curling of the spiral arm, and in the case of NGC\,7213, a small pitch angle ($p = 13^{\circ}$) is 
required to curl the spiral arm close to the inner radius and give relative more importance to the inner regions of the disk, 
and consequently, better reproduce the wings of the profile. Figure \ref{model_fit} shows the best fit for each epoch while the corresponding parameters are listed in Table \ref{model_pars}.
Surface emissivity maps corresponding to the models are shown in Figure \ref{disk_model}.




\begin{center}
\begin{deluxetable}{l c c c c}
\tablecolumns{5}
\tablecaption{\ngc{7213} best fit parameters of the accretion disk model}
\tablehead{UT Date		&MJD 	&$\xi_q$ 		 &A   	    &$\phi_0 (^\circ)$}
\startdata    
Sep 27 2011 & 55831.130  & 1800  &  4.0   &  +280\\    
Jul 21 2012 & 56129.155  & 2000  &  3.0   &  +205\\
Oct 15 2012 & 56215.111  & 2500  &  2.0   &  +45\\
Nov 22 2012 & 56253.047  & 2500  &  3.0   &  +25\\
Apr 13 2013 & 56395.383  & 2500  &  2.0   &  -30 \\
May 20 2013 & 56432.386  & 2500  &  1.8   &  -50 \\
Jun 14 2013 & 56457.399  & 3000  &  1.0   &  -60 \\
Jul 07 2013 & 56480.218  & 3000  &  0.2   &  -80\\ 
\label{model_pars}
\tablecomments{Column (1) gives the date of observations while column (2) gives the Modified Julian Date (JD$-2400000.5$).
Column (3) gives the break or saturation radius $\xi_q$, column (4) gives the contrast of the spial arm $A$ and column (5) gives the orientation of the spiral arm $\phi_0$ as it rotates in the disk.}
\end{deluxetable}
\end{center}

\subsection{The central broad component (CBC)}
\label{cbc}

%

Figure \ref{model_fit} shows that the accretion disk model can reproduce the broad double-peaked emission line in our observations, and we thus confirm that its emission originates from gas rotating in a relativistic Keplerian disk with very high projected velocities (e.g., $\sim$ 8,550 \kms\ is the velocity separation between the blue and red peaks of the double-peaked profile of 2011 September 27).

But the $rms$ spectrum of Figure \ref{rms_only} shows that, besides the broad double-peaked component there is another varying component that appears as an excess variation above that attributed to the gas that is rotating in the accretion disk, observed at lower project velocities. We have called this component CBC (Central Broad Component).
The CBC is not easily seen in the profile because of the blended narrow emission lines H$\alpha$+[NII] that sit ``on top" of it. It cannot be due to the narrow lines because the observed variation is larger than the estimated 8\% variations attributed to the uncertainties in the intercalibration using the narrow lines. 

We inspected the IFU data in order to check if the CBC and the broad double-peaked line were 
detectable outside the nucleus, considering our spatial resolution of 58 pc. 
We did not find evidence for the presence of the CBC neither the double-peaked line beyond this 58 pc region, 
concluding that both of them are unresolved, consistent with an origin in the outer parts of the accretion disk or in the BLR.

\begin{figure*}
 \includegraphics[width=0.5\textwidth]{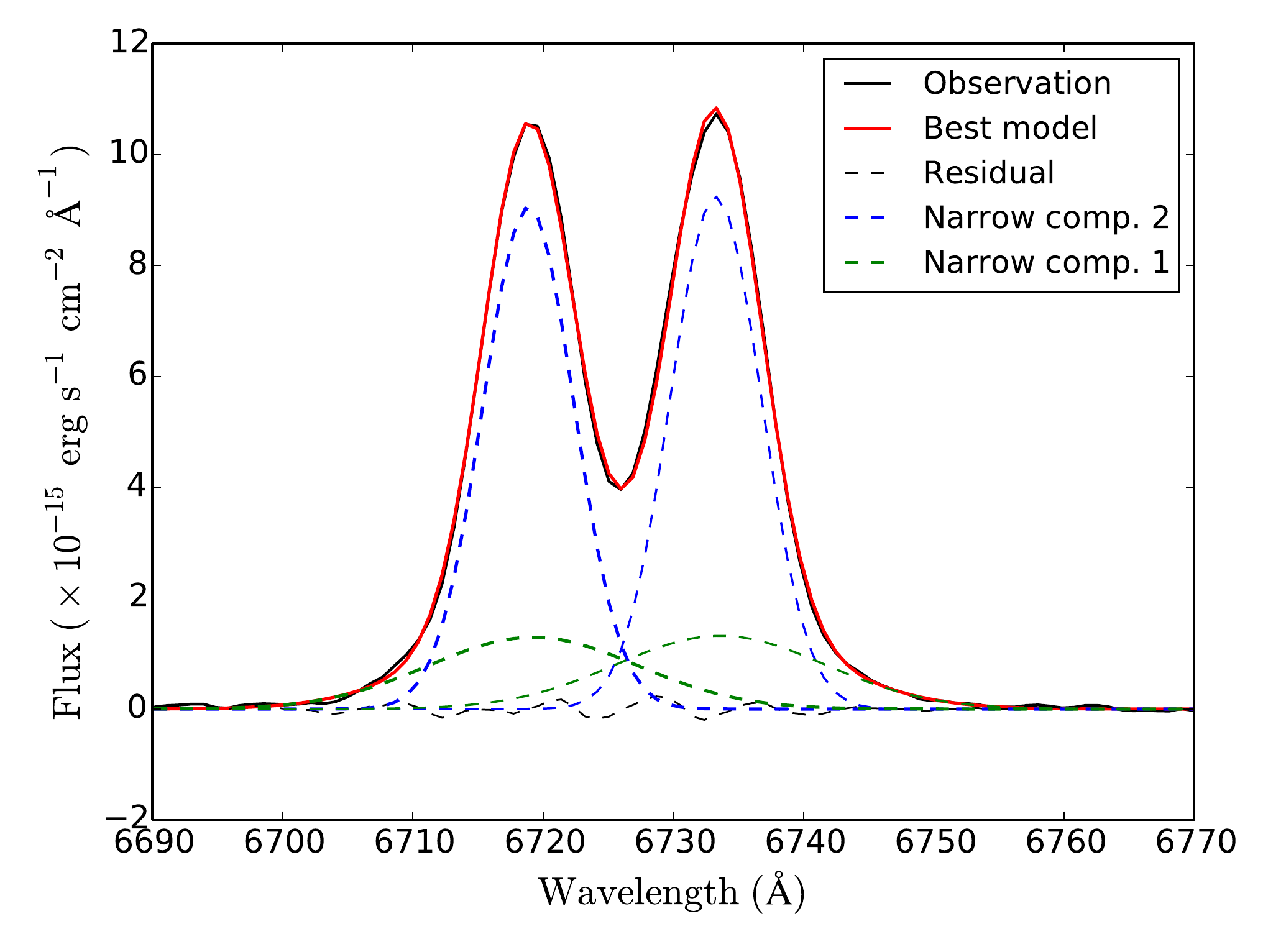}
 \includegraphics[width=0.5\textwidth]{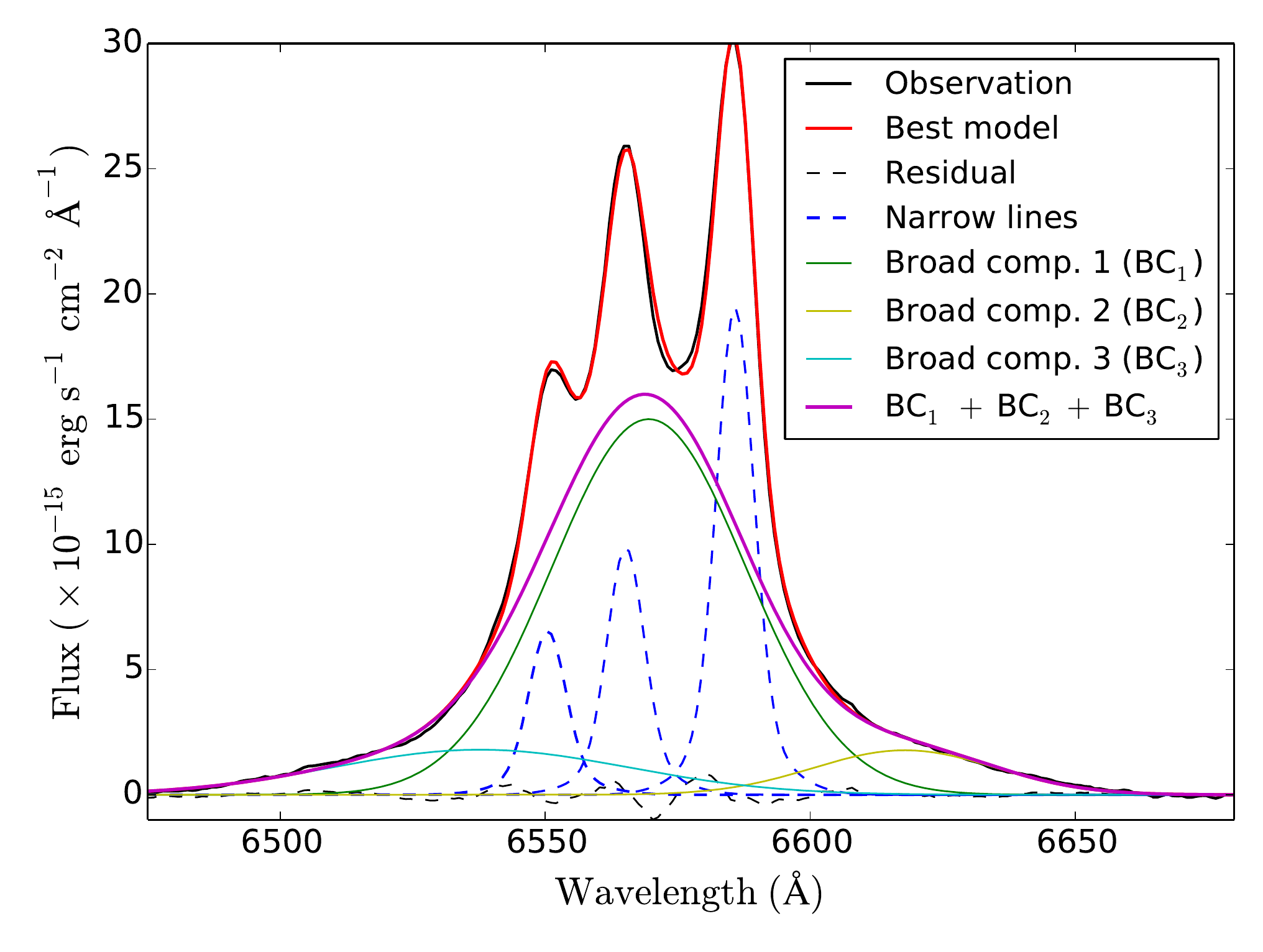}
 \caption{\emph{Left}: Fit of the narrow emission lines [SII] $\lambda \lambda$6716, 6731. Each narrow line was fitted using two gaussian components which are represented by the blue and green dashed lines. The central velocity and velocity width of each component was constrained to be the same for both [SII] lines. The red solid line is the best fit to the lines and represent the sum of the individual components, and the black dashed line is the residual from the fit.
 \emph{Right}: Fit of the H$\alpha$ CBC and narrow components of H$\alpha$+[NII]\,$\lambda \lambda$6548, 6583. The narrow components were constrained to have the same velocities as the [SII] lines. The fitted components to the narrow lines are shown as dashed blue lines. The CBC was fitted with three Gaussian components; the sum of the three is 
shown as a continuous magenta line and the individual components are shown in green, yellow and cyan.The solid red line is the sum of all components and the residual (observed - modeled) is shown as a black dashed line.}
 \label{sii_fit}
 \end{figure*}

In order to measure the contribution of the CBC
in each epoch and characterize its profile, it is necessary to separate the CBC from the narrow lines. We have used the following method to do this:

\begin{itemize}

 \item We first subtracted the modeled double-peaked profile from each spectrum, and then fitted the narrow emission lines [SII] $\lambda \lambda$6716, 6731. These lines were fitted with two Gaussian components each, as the  use of only one Gaussian did not fit well the base of the lines. Each Gaussian component was constrained in velocity space to have the same central velocity and width in the two lines of the [SII] doublet. An example of the fit to the [SII] lines is shown in the left pannel of Figure \ref{sii_fit}.
 The broader of the two narrow components of [SII] $\lambda \lambda$6716, 6731 has typical central velocity of 66$\pm$17\,km\,s$^{-1}$ and velocity dispersion of 329$\pm$17\,km\,s$^{-1}$;
 the other narrow component has central velocity and velocity dispersion of 41$\pm$8 and 159$\pm$5\,km\,s$^{-1}$, respectively.
 
 \item Since the narrow emission line components of  H$\alpha$+[NII]\,$\lambda \lambda$6548, 6583 are superposed on the H$\alpha$ CBC that we want to characterize, in order to decrease the degeneracy in the fit, we adopted the physically motivated assumption that these components originate in the same region as the [SII] lines, thus fixing the centroid velocities and widths of the Gaussians to the corresponding values of the [SII] lines. The flux of the lines was allowed to vary, only constraining the flux of [NII]\,$\lambda$6583 to be 2.87 times the flux of [NII]\,$\lambda$ 6548 \citep{Osterbrock}.

 \item Once we constrained the center and dispersion velocities of the H$\alpha$+[NII] narrow components, (and the flux constraints above) we fitted the H$\alpha$ CBC together with these lines. For the fit of the CBC we used three Gaussian components. An illustration of the result of this fit is shown in the right panel of Figure \ref{sii_fit}. The magenta line 
represents our fit for the CBC, whose flux accounts for a significant part of the total broad H$\alpha$ line (comprising the contributions of the double-peaked component plus the CBC).

 \end{itemize}

After fitting the broad double-peaked H$\alpha$, the narrow emission lines H$\alpha$+[NII]$\lambda\lambda$6548,83 and the CBC, we calculated the \emph{mean}, \emph{rms} and \emph{minimum} spectra from our 8 modeled observations,
as well as of the modeled profiles.

\begin{figure}
\centering
\includegraphics[width=8cm]{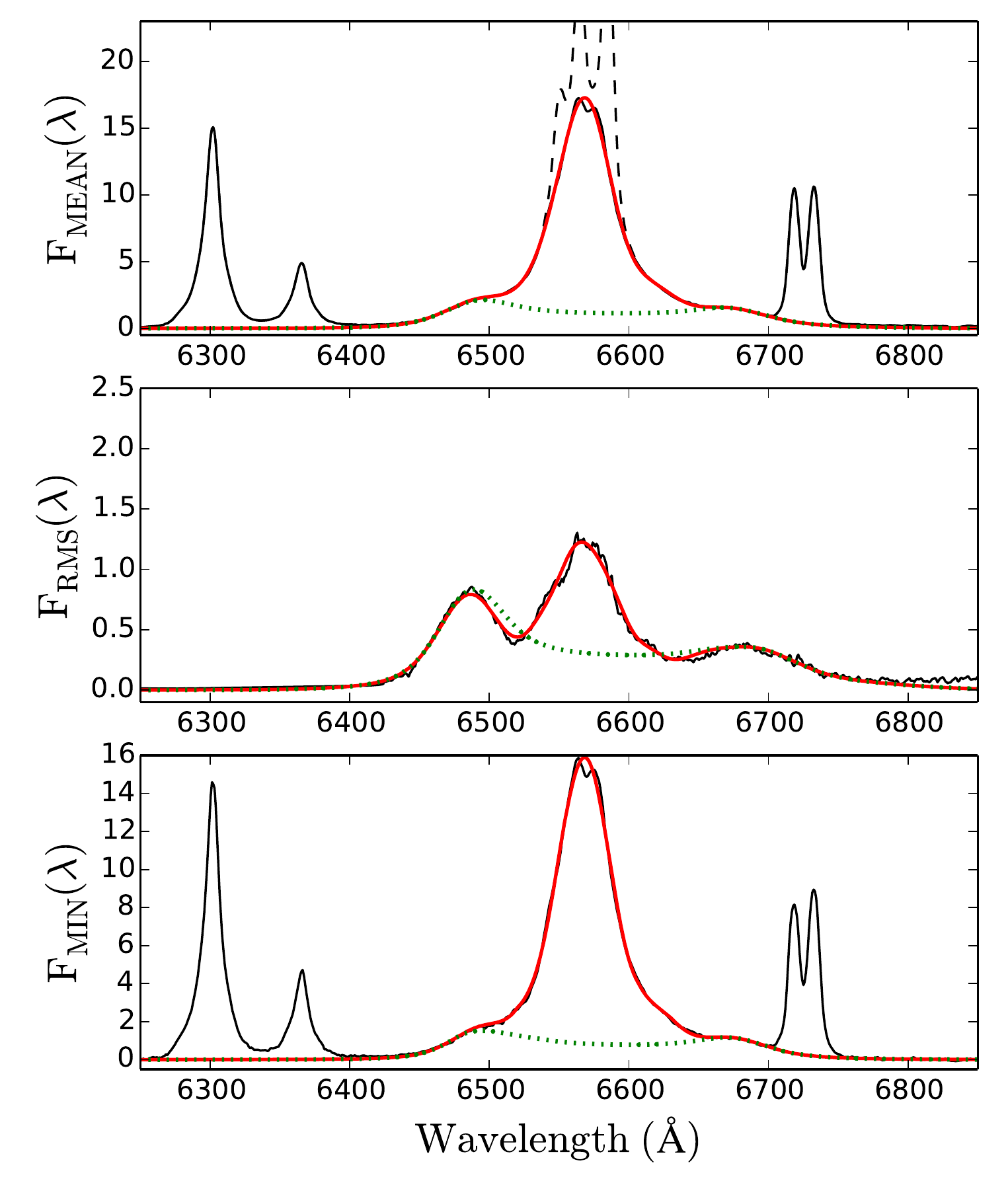}

\caption{\emph{Top}:
The black dashed line is the mean spectrum calculated from our observations, 
while the solid black line is the mean spectrum calculated after subtracting the contribution
of the H$\alpha$+[NII]\,$\lambda \lambda$6548, 6583 narrow lines.
The green dotted line is the mean model spectrum of the accretion disk.
The solid red line is the mean spectrum of the total broad H$\alpha$ emission
calculated from the accretion disk models + the fitted CBC.
\emph{Middle}: the $rms$ variations of the observed and modeled spectra.
\emph{Bottom}: the minimum observed and modeled spectra.
}
\label{diff_mean}
\end{figure}

The top pannel of Figure \ref{diff_mean} shows the mean spectrum we have calculated from our data.
This spectrum displays double-shoulders, with the maximum flux density of the blue side of 
the profile slightly higher than that of the red side, which is well modelled by our mean accretion disk model,
shown as the green dotted line.
In addition we note that the CBC $H\alpha$ component also contributes strongly to the mean
overall broad $H\alpha$ emission.

Before calculating the $rms$ spectrum we subtracted the contribution of the narrow emission lines, and the 
result is shown in the middle pannel of Figure \ref{diff_mean}.
The $rms$ spectrum, for both the data and the models (red line), shows that, as discused before, the most variable portion of the spectrum shows a profile with three peaks: a blue peak centered around 6487\,\AA, a red peak centered around 6684\,\AA, in agreement with the expected variations in the accretion disk plus a central peak corresponding to variations in the CBC. The similarity between the black (observations) and red (model) profiles shows that the variations in the broad profile can be well reproduced by the $rms$ profile of our accretion disk model (green dotted line) plus the CBC (modelled  via the fit of three gaussinas to each profile).

The minimum spectrum was constructed by selecting, at each wavelength, the minimum flux
from all spectra. This minimum spectrum represents a base profile
which is common to all profiles, and is shown at the bottom pannel of Figure \ref{diff_mean}. 
In the scenario in which
the emission arises from a circular accretion disk with an emissivity enhancement (such as a spiral arm,
as we have considered in our modelling),
the minimum spectrum would be that of the underlying accretion disk.
Figure \ref{diff_mean} shows that
the minimum spectrum also displays two shoulders, consistent with the above assumption. 
Individually or taken together, the mean, $rms$ and minimum spectra all support the scenario where 
the origin of the broad, double-peaked line profile is indeed a rotating disk. 
The CBC is evidence of additional gas at lower line-of-sight velocities than the gas in the accretion disk, probably orbiting the SMBH at distances beyond the disk.

\subsubsection{Properties of the CBC}

We have mesured three properties of the CBC:
the integrated flux, \fcbc, the peak velocity,
V$_{\text{peak}}$, which is the velocity corresponding to the wavelength
of maximum flux and the median velocity, V$_{50}$, which corresponds to
the wavelength at which the integrated flux under the profile corresponds to 50\% of the total flux.
As the profile is not fitted by only one Gaussian and contain asymmetries, we adopted as velocity dispersion the velocities above and below the median velocity that encompass $34\%$ of the flux under the profile,  $\pm\sigma_V$.
We list these measurements in Table \ref{vbroad_model}.

We note that the CBC flux \fcbc,
is usually $\sim$\,$50 --100\%$ higher than the  flux of the double-peaked component, \fdp\ (the only exception being the first epoch).
However, during the campaign, the amplitude of the variations of the \fcbc\ was lower than that of the \fdp: while 
\fdp\ varied between minimum and maximum fluxes of 410 and 802\,$\times10^{-15}$\,erg\,s$^{-1}$\,cm$^{-2}$, respectively, \fcbc\ varied between minimum and maximum fluxes of 
803 and 921\,$\times10^{-15}$\,erg\,s$^{-1}$\,cm$^{-2}$, respectively.
This behavior is also evident from a comparison of the top and middle pannels of Figure \ref{diff_mean}: the CBC is much more pronounced in the mean
spectrum than in the rms spectrum.
We define a proxy for the velocity width of the CBC as:
\begin{equation}
W_{68} = +\sigma_V - (-\sigma_V)
\end{equation}
\noindent which represents the velocity width of the line that contains 68\%\ of the
total line flux around V$_{50}$.
The values of W$_{68}$ are listed in the last column of Table \ref{vbroad_model}. 
The average value of the parameter W$_{68}$ is $2100\pm73$ \kms\ which represents
the average velocity width of the CBC.

\begin{center}
\begin{deluxetable}{l c c c c c}
\tablecolumns{6}
\tablecaption{\ngc{7213} CBC properties}
\tablehead{UT Date  &MJD    &\fcbc &V$_\text{peak}$         &V$_{50}\pm \sigma_{V}$ &W$_{68}$}
\startdata
 Sep 27 2011     &55831.130 &921$\pm$74 &253 &274$^{+1046}_{-920}$  & 1966\\ 
 Jul 21 2012     &56129.155 &936$\pm$75 &211 &148$^{+1004}_{-1045}$ & 2049\\ 
 Oct 15 2012     &56215.111 &891$\pm$71 &253 &274$^{+1088}_{-1087}$ & 2175\\ 
 Nov 22 2012     &56253.047 &803$\pm$64 &253 &274$^{+1046}_{-1003}$ & 2049\\ 
 Apr 13 2013     &56395.383 &829$\pm$66 &253 &274$^{+1046}_{-1003}$ & 2049\\ 
 May 20 2013     &56432.315 &831$\pm$67 &253 &274$^{+1130}_{-1003}$ & 2133\\ 
 Jun 14 2013     &56457.399 &889$\pm$71 &253 &274$^{+1171}_{-1003}$ & 2174\\ 
 Jul 07 2013     &56480.218 &904$\pm$72 &253 &274$^{+1171}_{-1003}$ & 2174\\ 
\label{vbroad_model}
\tablecomments{Column (1): date of observations; 
column (2): Modified Julian Date (JD$-2400000.5$);  
column (3): integrated flux of the CBC in units of $10^{-15}$\,erg\,s$^{-1}$\,cm$^{-2}$ and the uncertainties are $\pm$\,8$\%$ of the measured flux;
column (4): peak velocity of the line; 
column (5): median velocity of the line plus/minus the velocities corresponding to $\pm34\%$ of the integrated flux;
column (6): velocity width of the CBC that contains 68\% of the total line flux around V$_{50}$ . Velocity units are \kms.}
\end{deluxetable}
\end{center}

\section{Discussion}
\subsection{Low-state of the double-peaked profile}
Differently from what was observed by \citet{Schimoia15} for the \emph{low state} of the double-peaked profile of NGC\,1097, when NGC\,7213 is in a low state it is hard to detect the asymmetry in the broad profile. 
This means that, when the \fdp\ flux is low, the contrast between the spiral arm and the underlying disk fades.
This fading can be observed in the emissivity maps of Figure \ref{disk_model}.
In the first epoch the contrast parameter was A=4.0 but in the last it was very small: A=0.2. 
Between 2013 April 13 (MJD\,5639) and 2013 July 23 (MJD\,56496), the 
overall flux \fdp\ stayed low (did not show much variation),  and the ratio \fratio\ stayed almost the same, around 1.5.
This result is in agreement with the decrease in the contrast between the arm and the underlying disk during the
last observations.




\subsection{Implications for the mass of the SMBH}
The epoch that most clearly shows the blue and red peaks in the double-peaked profile is 2011 September 27.
In this profile we measure a velocity separation between the blue and red peaks of $\Delta v \approx 8850$ \kms. 
Considering half of this velocity as a characteristic velocity for the gas in the disk, corresponding approximately  to the  radius of maximum emission, and correcting this value for the inclination of the disk, 
 we obtain a circular velocity $V_C = \frac{8850}{2}\times\frac{1}{\sin(47^{\circ})}\sim6050$\,\kms.
%

As discussed in \S \ref{results}, the shortest variability timescale in \fdp\ probably lies between 7 -- 28 days. If we adopt 17 -- the middle between these two limits -- as the light-travel time between the central ionizing source (that we adopt as approximately coinciding with the location of the SMBH) and the radius of maximum emission,  we can estimate the mass of the SMBH as \mbh{1.21}{8}.
Considering that this radius could be in the range between the lower and upper limits of 7 and 28 days, respectively, the resulting range for the SMBH mass estimate is 
5$\times10^{7}<\mathrm{M_{\bullet}}<2\times10^8$\,\msun.


An independent estimate of the mass of the SMBH can be obtained via the $M_{\bullet} - \sigma_{\star}$ relation of \citet{Tremaine02}:
\begin{equation}
\label{msigma}
 \log \left ( \frac{M_{\bullet}}{M_{\odot}} \right )= \alpha + \beta \log \left ( \frac{\sigma_{\star}}{\sigma_0}\right )
\end{equation}

\noindent where $\alpha=8.13\pm0.06$; $\beta=4.02\pm0.32$; and $\sigma_0 = 200$\,\kms. The scatter in the \msigma\ 
relation is not included in the error, but the error bars on the coefficients are.

When we performed the stellar population synthesis with {\tt starlight-v04} we used the simple stellar population templates of \citet{Bruzual03}
and allowed the fit of the stellar kinematics. The value we obtained for the velocity dispersion was $\sigma=219$\,\kms. 
We have corrected this value for the 
instrumental resolution, $\sigma_{inst}$, and the resolution of the template spectra, $\sigma_{base}$, as follows: 

\begin{equation}
\sigma_{\star}^{2} = \sigma^{2} - \sigma_{inst}^{2} + \sigma_{base}^{2}
\end{equation}

The B600 grating that we used in the longslit GMOS observations has an instrumental resolution of $\sigma_{inst}=177$\,\kms, while the \citet{Bruzual03} 
base of simple stellar population spectra has a resolution of R $=2000$, thus  $\sigma_{base}=150$\,\kms, in the wavelength range 3200 \AA\ to 9500\AA.
We thus obtain a stellar velocity dispersion of $\sigma_{\star} = 198$\,\kms, which via Equation \ref{msigma} gives \mbh{1.29}{8}, 
in very good agreement with the value we have obtained via the estimated light travel time between
the nuclear ionizing source and the radius of maximum emission of the line emitting disk.

This value that we have obtained for the stellar velocity dispersion is also very close to the value $\sigma_{\star}=185$\,\kms\ 
previously obtained by \citet{Woo02}, who also used the $M_{\bullet} - \sigma_{\star}$ relation \citep{Tremaine02} to estimate the
mass of the SMBH as \mbh{9.77}{7}. Considering that there is an intrinsic scatter of 
$\sim0.3\,\mathrm{dex}$ (a factor $\sim$ 2) in the $M_{\bullet} - \sigma_{\star}$ relation, we conclude that both 
our determinations of the mass of the SMBH are in good agreement with this previous determination, supporting this value.

\subsection{Variability timescales}

The two shortest variability timescales of standard accretion disks are the \emph{light travel timescale}, $\tau_l$, and
the \emph{dynamical timescale}, $\tau_{dyn}$ \citep{Frank02}:
\begin{equation}
 \tau_l = 6 M_8 \xi_3\ \text{days}
\end{equation}
\begin{equation}
 \tau_{dyn} = 6 M_8 {\xi_3}^{3/2}\ \text{months}
\end{equation}

\noindent where $M_8$ is the mass of the SMBH in units of 10$^8$\msun\ and $\xi_3 = \xi \times 10^{-3}$. Adopting the mass of the SMBH in NGC\,7213 as \mbh{1.29}{8} and 
considering the best model inner and outer radius of the accretion disk, $0.3<\xi_3<3$, we obtain a range for the variability timescales of:
\begin{itemize}
 \item $\tau_l$: 2.3 -- 23 days
 \item $\tau_{dyn}$: 1.3 -- 40 months
\end{itemize}

\begin{figure}
\centering
\includegraphics[width=8cm]{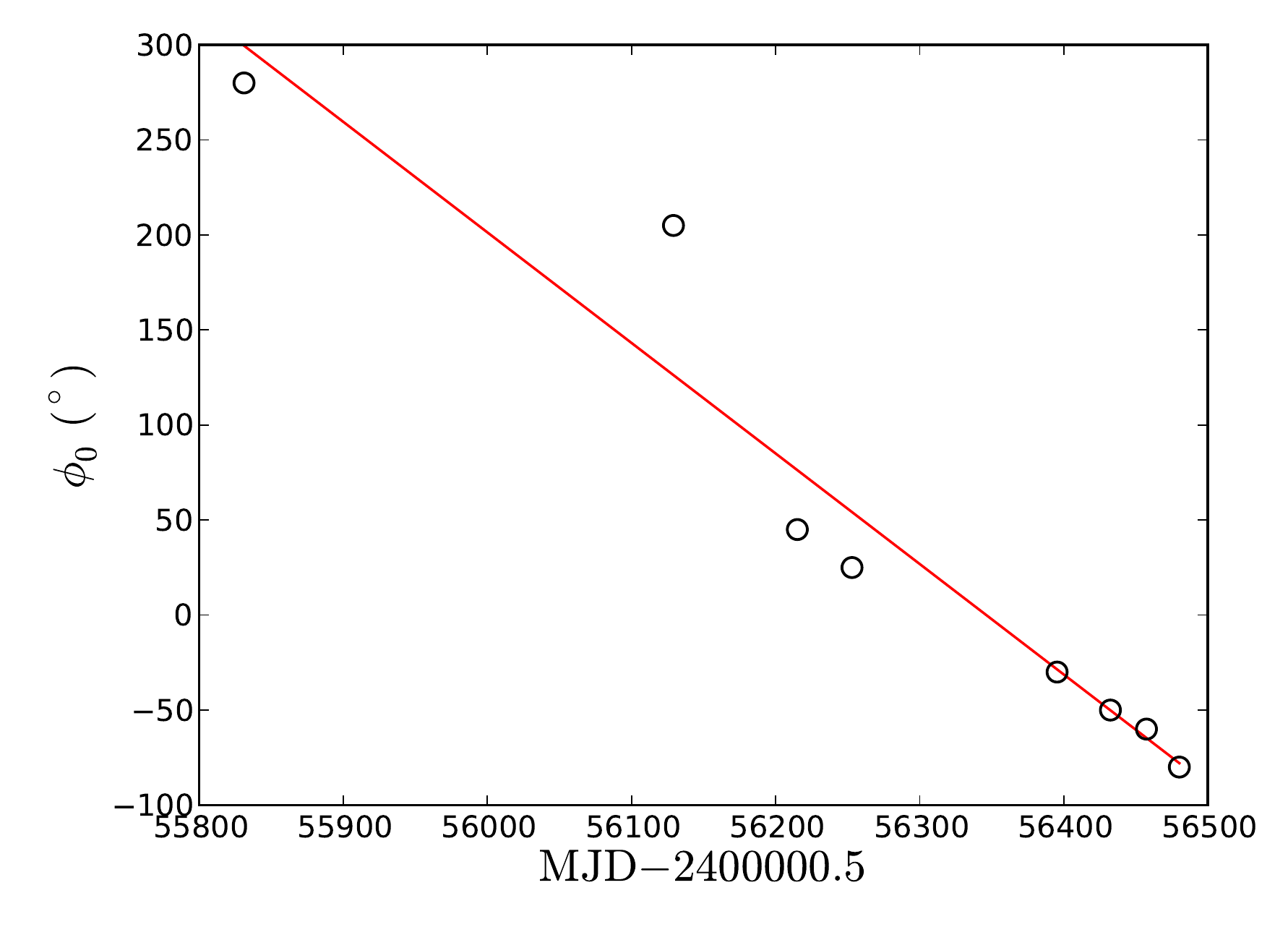}
\caption{The black open circles represent the parameter $\phi_0$ for each modeled epoch. The red solid line is the best linear fit to the data, which has an angular
coefficient of $\alpha\approx-0.58^{\circ}$\,day$^{-1}$. The rotation period corresponds to $P\approx 620$ days or 21 months.}
\label{phi0}
\end{figure}

The evolution of the model parameter $\phi_0$ -- the azimuthal angle of the spiral pattern, with time is shown in Figure \ref{phi0}. 
A  linear fit to the data gives an angular velocity of $\alpha\approx-0.58^{\circ}$\,day$^{-1}$, 
which implies a rotation period of $\sim$ 620 days or 21 months, within the expected range of the dynamical timescale above. 
Considering this period of rotation as a dynamical timescale of the accretion disk, it suggests that the asymmetric feature of the accretion disk has completed almost 
one full rotation. The observed precession period of the spiral pattern is also in good agreement with the theoretical predictions for m=1
modes in massive, Keplerian disks by \citet{Adams89} and \citet{Shu90}.

From Figure \ref{light_curve} it can be further concluded that variations in the relative intensity of fluxes of the red 
and blue sides of the profile, \fratio, occur in the
timescale of a few months. For instance: \fratio\ varied from 1.09 (2011 September 27) to 1.33 (2012 July 21) 
in almost 10 months and also varied from 1.33 (2012 July 21) to 1.67 (2012 November 22) in approximately 4 months. This variation is also within the range of the dynamical timescale, and compatible with the rotation period of the spiral arm in the accretion disk discussed above. 

Regarding the shortest variability timescale, although our relatively sparse
monitoring did not allow us to put strong constraints on its value,
the range we have estimated, between 7 and 28 days is approximately consistent with the light travel time.


%

\subsection{The CBC}



The mean velocity width of the CBC, $\overline{\mathrm{W}_{68}} = 2100\,\pm73$\,\kms\, suggests it is located at larger distances from the SMBH than the disk. 
Estimating its variability timescale from the line flux -- considering typical uncertainties of $\sim$ 8\% -- we obtain
$\sim\,120$ days, $\sim$ 6 times the outer radius of the disk. The central velocity of the CBC of V$_{50}=260\,\pm\,40$\,\kms\ is much larger than the gravitational redshift at the above distance, that suggests
the presence of a bulk motion of this region.

The presence of a CBC together with disk-like profiles have been also found by other authors \citet{Ho97} and \citet{SB16}.
Their methodology for subtracting the contribution of the narrow lines and  fitting the broad central component 
is very similar to the methodology adopted in the present work. These \textbf{authors} found mean values of the FWHM for the broad component
of $\sim2000$\,\kms\ \citep{Ho97} and $\sim1500\pm500$\ \citep{SB16}, which are consistent with the value of $\overline{\mathrm{W}_{68}} = 2100\,\pm73$\,\kms\ we found for NGC\,7213.

\section{Conclusions}
We have presented 13 new optical spectra of the broad H$\alpha$ profile 
of the AGN in NGC\,7213 over a time span of 22 months, with observations sparsed by time intervals from a week to a few months. The main results of this paper are:

\begin{itemize}

\item It is the first time that variability is reported for the broad ($\approx 8550$\,\kms for the velocity separation of the blue and red peaks) H$\alpha$ line of this AGN, that shows a double-peaked profile, typical of gas rotating in an accretion disk around a SMBH;

\item The relative intensity of the integrated flux of the blue and red sides of the double-peaked profile displayed significant variations on a timescale $\gtrsim 3$ months, consistent with the \emph{dynamical timescale} of gas rotating in an accretion disk around a 10$^{8}$\,M$_\odot$ SMBH;

 \item The total flux of the broad line showed variations on a timescale between 7 and 28 days, consistent with the  \emph{light travel timescale} between the ionizing source and the emitting part of the disk;

 \item The $rms$ variation specrum reveals that the most variable part of the broad H$\alpha$ line shows three peaks: a blue and a red peak consistent with an origin in the accretion disk plus an additinal central peak which we attribute to a ``central broad component" (CBC), showing similar amplitude to those of the two other peaks;

\item We successfully modeled the broad double-peaked profile as due to gas emission from a region of a Keplerian and relativistic accretion disk with inner and outer radii of $\xi_1=300\pm60$ and $\xi_2=3000\pm90$ (in units of gravitational radii), respectively, and inclination angle of $i=47^{\circ}\pm2^{\circ}$ relative to the plane of the sky. We also found that the disk harbors a spiral arm with varying contrast relative to the underlying disk;
 
\item The variations in the relative intensity of the blue and red sides of the profile were modeled as due the rotation of the spiral arm, with a period of $\sim$ 21 months. This arm completed almost one full rotation in the accretion disk and faded (decreased its contrast) over the approximately two years spanned by the observations;

\item The profile of the CBC was modelled via the fit of three Gaussians; it shows an average width velocity (proxy for the velocity dispersion) of W$_{68} = 2100\pm73$\kms,
which suggests that the gas of this component is located at larger distances from the SMBH than the outer radius of the disk;

 

 \item Using the fit of the stellar absorption features with the {\tt starlight-v04} code we obtained a velocity dispersion of the bulge 
of $\sigma_{\star} = 188$\kms, what implies a mass for the SMBH of \mbh{1.29}{8}\ via the $M_{\bullet} - \sigma_{\star}$ relation, in agreement with previous determinations;

 \item We also estimated the mass of the SMBH 
using a representative velocity of the gas in the disk and the light travel time between the SMBH and the disk. We find that the SMBH mass is in the range $5\times10^{7}<\mathrm{M_{\bullet}}<2\times10^8$\,\msun, also in agreement with the previous and our above estimate.
\end{itemize}


In summary, our proposed scenario for the origin of the broad H$\alpha$ profile of NGC\,7213 is the following: the broad double-peaked emission arises from a Keplerian and relativistic accretion disk, inclined by 47 degrees relative to the plane of the sky from a region with inner and outer radii of $\approx$\,300 and 3000 gravitional radii, respectively.
This disk orbits a SMBH with a mass in the range $5\times10^7\leq$\,M$_{\bullet}$\,$\leq 2\times10^8$\,\msun.
The relative intensity of the flux of blue and red sides of the double-peaked profile changes due to the rotation 
of a spiral arm with a rotation period of 21 months. The contrast between the the arm and underlying disk
decreased gradually during the approximately two years of observations, leading to ever smaller asymmetries between the heights of the blue and red sides of the double-peaked profile. An additional component, the CBC, also contributes to the variable broad H$\alpha$ profile as a central ``hump" observed at velocities close to systemic, with a velocity width of $2100$\,\kms. 
We propose that this component originates in gas that is also orbiting the black hole but is either at larger radii in the accretion disk or not coplanar with the disk.
\acknowledgments
J.S.S. acknowledges CNPq, National Council for Scientific and Technological Development - Brazil.
R.N. acknowledges support from FAPESP.


\begin{thebibliography}{}

\bibitem[\protect\citeauthoryear{Adams, Ruden \& Shu}{1989}]{Adams89}  
Adams F.~C., Ruden S.~P., Shu F.~H., 1989, ApJ,  347, 959 


\bibitem[\protect\citeauthoryear{Bianchi et al.}{2003}]{Bianchi03}  
Bianchi S., Matt G., Balestra I., Perola G.~C., 2003,   
\textsl{The origin of the iron lines in NGC 7213},  
A\&A,  407, L21 


\bibitem[\protect\citeauthoryear{Bianchi et al.}{2008}]{Bianchi08}  
Bianchi S., La Franca F., Matt G., Guainazzi M., Jimenez Bail{\'o}n E., 
Longinotti A.~L., Nicastro F., Pentericci L., 2008,   
MNRAS,  389, L52


\bibitem[\protect\citeauthoryear{Bruzual \& Charlot}{2003}]{Bruzual03}  
Bruzual G., Charlot S., 2003,   
MNRAS,  344, 1000 


\bibitem[\protect\citeauthoryear{Chen et al.}{1989}]{Chen89}  
Chen K., Halpern J.~P., \& Filippenko A.~V., 1989,   
ApJ,  339, 742 


\bibitem[\protect\citeauthoryear{Chen \& Halpern}{1989}]{CeH89}  
Chen K., \& Halpern J.~P., 1989,   
ApJ,  344, 115


\bibitem[Cid Fernandes et al.(2005)]{Cid}  
Cid Fernandes R., Mateus A., Sodr{\'e} L., Stasi{\'n}ska G., \& Gomes J.~M., 
2005,   
MNRAS,  358, 363 



\bibitem[\protect\citeauthoryear{Dumont \& Collin-Souffrin}{1990}]{Dumont90}  
Dumont A.~M., Collin-Souffrin S., 1990,   
A\&A,  229, 313

\bibitem[\protect\citeauthoryear{Emmanoulopoulos et al.}{2012}]{Emmanoulopoulos12}  
Emmanoulopoulos D., Papadakis I.~E., McHardy I.~M., Ar{\'e}valo P., Calvelo 
D.~E., Uttley P., 2012,   
MNRAS,  424, 1327 




\bibitem[\protect\citeauthoryear{Eracleous \& Halpern}{2003}]{Eracleous03}  
Eracleous M., \& Halpern J.~P., 2003, 
ApJ,  599, 886 




\bibitem[\protect\citeauthoryear{Eracleous et al.}{2009}]{Eracleous09}  
Eracleous M., Lewis K. T. \& Flohic H. M. L. G., 2009, New Astr. Rev. 53, 133


\bibitem[\protect\citeauthoryear{Filippenko \& Halpern}{1984}]{Filippenko84}  
Filippenko A.~V., Halpern J.~P., 1984,   
ApJ,  285, 458 

\bibitem[\protect\citeauthoryear{Frank et al.}{2002}]{Frank02}  
Frank J., King A., Raine D.~J., 2002,   
apa..book,  398 

\bibitem[\protect\citeauthoryear{Gezari et al.}{2007}]{Gezari07}  
Gezari S., Halpern J.~P., Eracleous M., 2007,   
ApJS,  169, 167 

\bibitem[\protect\citeauthoryear{Gilbert et al.}{1999}]{Gilbert99}  
Gilbert A.~M., Eracleous M., Filippenko A.~V., \& Halpern J.~P., 1999,   
ASPC,  175, 189 


\bibitem[\protect\citeauthoryear{Ho et al.}{1997}]{Ho97}  
Ho L.~C., Filippenko A.~V., Sargent W.~L.~W., Peng C.~Y., 1997,   
ApJS,  112, 391 


\bibitem[\protect\citeauthoryear{Lewis et al.}{2010}]{Lewis10}  
Lewis K.~T., Eracleous M., \& Storchi-Bergmann T., 2010,   
ApJS,  187, 416

\bibitem[\protect\citeauthoryear{Lobban et al.}{2010}]{Lobban10}  
Lobban A.~P., Reeves J.~N., Porquet D., Braito V., Markowitz A., Miller L., Turner T.~J., 2010,   
MNRAS,  408, 551 




\bibitem[\protect\citeauthoryear{Narayan \& McClintock}{2008}]{Narayan08}  
Narayan R., \& McClintock J.~E., 2008,   
NewAR,  51, 733 




\bibitem[\protect\citeauthoryear{Osterbrock \& Ferland}{2006}]{Osterbrock}  
Osterbrock D.~E., Ferland G.~J., 2006,   
\textsl{Astrophysics of gaseous nebulae and active galactic nuclei},  agna.book, 


\bibitem[\protect\citeauthoryear{Phillips}{1979}]{Phillips79}  
Phillips M.~M., 1979,   
ApJ,  227, L121 

\bibitem[\protect\citeauthoryear{Schimoia et al.}{2012}]{Schimoia12}  
Schimoia J.~S., Storchi-Bergmann T., Nemmen R.~S., Winge C., \& Eracleous M., 2012,   
ApJ,  748, 145 

\bibitem[\protect\citeauthoryear{Schimoia et al.}{2015}]{Schimoia15}  
Schimoia J.~S., Storchi-Bergmann T., Grupe D., Eracleous M., Peterson B.~M., Baldwin J.~A., Nemmen R.~S., Winge C., 2015,   
ApJ,  800, 63 


\bibitem[\protect\citeauthoryear{Schnorr-M{\"u}ller et al.}{2014}]{Allan14}  
Schnorr-M{\"u}ller A., Storchi-Bergmann T., Nagar N.~M., Ferrari F., 2014,  
MNRAS,  438, 3322 

\bibitem[\protect\citeauthoryear{Shakura \& Sunyaev}{1973}]{Shakura}  
Shakura N.~I., \& Sunyaev R.~A., 1973,   
IAUS,  55, 155 


\bibitem[\protect\citeauthoryear{Shu et al.}{1990}]{Shu90}  
Shu F.~H., Tremaine S., Adams F.~C., Ruden S.~P., 1990,   
ApJ,  358, 495 



\bibitem[\protect\citeauthoryear{Storchi-Bergmann et al.}{1996}]{SB96}  
Storchi-Bergmann T., Rodriguez-Ardila A., Schmitt H.~R., Wilson A.~S., Baldwin J.~A., 1996,   
\textsl{Circumnuclear Star Formation in Active Galaxies},  
ApJ,  472, 83 


\bibitem[\protect\citeauthoryear{Storchi-Bergmann et al.}{2003}]{SB03}
Storchi-Bergmann T., Nemmen da Silva R., \& Eracleous M., et al., 2003,   
ApJ,  598, 956 


\bibitem[\protect\citeauthoryear{Storchi-Bergmann et al.}{2016}]{SB16}
Storchi-Bergmann T., Schimoia J.S.,  Peterson B.M., Elvis M., Denney K.D. , Eracleous M. \&  Nemmen da Silva R., 2016,
submitted 


\bibitem[\protect\citeauthoryear{Strateva et al.}{2003}]{Strateva03}  
Strateva I.~V., Strauss M.~A., Hao L., et al., 2003,   
AJ,  126, 1720


\bibitem[\protect\citeauthoryear{Tremaine et al.}{2002}]{Tremaine02}  
Tremaine S., Gebhardt K., Bender R., et al., 2002,   
\textsl{The Slope of the Black Hole Mass versus Velocity Dispersion 
Correlation},  
ApJ,  574, 740 


\bibitem[\protect\citeauthoryear{Woo et al.}{2002}]{Woo02}  
Woo J.-H., Urry C.~M., 2002,   
ApJ,  579, 530 


\bibitem[\protect\citeauthoryear{Yuan \& Narayan}{2014}]{Yuan14}  
Yuan F., Narayan R., 2014,   \textsl{Hot Accretion Flows Around Black Holes},  ARA\&A,  52, 529 




%
%
%
%
%

%

%

%
%
%
%
%
%

%
%
%
%

%
%
%
%
%
%

%
%
%
%
%
%
%
%
%
%
%
%
%
%
%
%



%
%
%
%
%

%
%
%
%
%

%
%
%



\end{thebibliography}
\end{document}